\documentclass[a4paper]{cas-dc}

\usepackage[numbers,sort&compress]{natbib}

\usepackage{enumitem}
\usepackage{url}
\usepackage{siunitx}
\usepackage{subcaption}

\def\tsc#1{\csdef{#1}{\textsc{\lowercase{#1}}\xspace}}
\tsc{WGM}
\tsc{QE}

\newcommand{\figref}[1]{Fig.\,\ref{#1}}
\newcommand{\Figref}[1]{Figure\,\ref{#1}}
\renewcommand{\tabref}[1]{Table\,\ref{#1}}
\newcommand{\eref}[1]{Eq.\,(\ref{#1})}
\newcommand{\secref}[1]{Sec.\,\ref{#1}}
\newcommand{\Secref}[1]{Section\,\ref{#1}}
\newcommand{\ie}{\emph{i.e.}}
\newcommand{\eg}{\emph{e.g.}}
\newcommand{\mE}{\mathbb{E}}

\ExplSyntaxOn
\cs_gset:Npn \__first_footerline: {}
\ExplSyntaxOff

\begin{document}
\let\WriteBookmarks\relax
\def\floatpagepagefraction{1}
\def\textpagefraction{.001}

\shorttitle{}    

\shortauthors{}  

\title [mode = title]{Optimal design of solar-battery hybrid resources considering multi-market participation under weather and price uncertainty}  

%

\author[1]{Hikaru Hoshino}[orcid=0000-0003-2487-8320]
\cormark[1]
\ead{hoshino@eng.u-hyogo.ac.jp}

\affiliation[1]{organization={University of Hyogo},
            addressline={2167 Shosha}, 
            city={Himeji},
            postcode={671-2280}, 
            state={Hyogo},
            country={Japan}}

\author[1]{Taiyo Mantani}
\ead{er24y025@guh.u-hyogo.ac.jp}

\author[1]{Eiko Furutani}[orcid=0000-0003-4344-8991]
\ead{furutani@eng.u-hyogo.ac.jp}


\cortext[1]{Corresponding author}



\begin{abstract}
The rapid growth of variable renewable energy has increased the need for flexible and efficiently coordinated energy resources. In this context, hybrid resources that combine renewable generation and battery storage within a single market-participating entity have attracted growing attention. Such hybrid resources can have multiple revenue streams, while allocating limited power and energy capacity across multiple electricity markets including energy and ancillary services. This multi-market coordination increases operational complexity and complicates profitability assessment, making optimal system sizing a challenging design problem. In addition, uncertainty in renewable generation and market prices makes it difficult for conventional optimization approaches to determine system designs that remain effective under stochastic operating conditions. To address these challenges, this paper proposes a deep reinforcement learning-based co-optimization framework for hybrid solar-battery resources. The framework embeds system design variables directly into the policy learning process, enabling joint optimization of hybrid system sizing and coordinated multi-market bidding strategies within a unified stochastic formulation. Case studies using historical renewable generation and market data demonstrate the effectiveness of the proposed framework  in identifying economically rational hybrid system design considering multi-market operation.
\end{abstract}





\maketitle
\section{Introduction}
\label{sec:introduction}

Hybrid resources~\cite{Ahlstrom2021}, which integrate multiple technologies into a single electricity-market entity, have rapidly gained attention as a powerful grid transformation approach. 
Unlike merely co-located resources that operate as separate market participants, hybrid resources are represented as a unified resource whose internal operation is coordinated across technologies. 
This configuration enables resource operators to jointly optimize generation, storage, and market participation decisions rather than relying on system operators to dispatch individual technologies independently~\cite{Ahlstrom2021}. 
While such integration offers significant physical, operational, and financial advantages, it also introduces additional operational and design complexity.

In this context, the operational strategies and physical system configuration of solar-battery hybrid resources are intrinsically coupled. 
The installed capacities of photovoltaic (PV) arrays and battery storage determine the feasible operational space across multiple electricity markets, while the expected revenues from these markets determine whether the upfront investment costs can be economically justified or not.  
At the same time, hybrid resource operators must dynamically allocate limited power and energy capacity across multiple markets under uncertainty in solar generation and market prices, while respecting intertemporal battery dynamics and inverter-limited dispatch constraints. 
As a result, determining an optimal hybrid resource configuration constitutes a challenging multi-scale decision problem that couples long-term investment decisions with short-term operational optimization under uncertainty.

Motivated by these challenges, this paper proposes a Deep Reinforcement Learning (DRL)-based co-optimization framework for PV-battery hybrid resources that integrates system design, multi-market bidding across energy and Ancillary Service (AS) markets, and battery control within a unified learning architecture. 
Through this integrated formulation, economically rational hybrid system configurations and coordinated bidding strategies can be learned directly under realistic operational and market conditions.

\subsection{Related Work}

The literature on renewable-battery hybrid resources has expanded rapidly in recent years, driven by the growing deployment of PV-battery and wind-battery systems as well as evolving market rules for their grid integration.
The existing work can be broadly categorized as follows:
\begin{enumerate}[label=\alph*)]
    \item \emph{Operational optimization of hybrid resources}: 
    A large body of research optimizes the operation of hybrid or co-located resources under the assumption that their physical configuration is fixed. 
    Early studies primarily focused on standalone (off-grid) systems or on participation limited to single energy markets, as reviewed in \cite{Olatomiwa2016}.
    More recent studies have shifted toward multi-market participation in which resources can exploit multiple revenue streams. 
    For example, in \cite{He2016}, optimal bidding strategies were studied for concentrating solar power plants participating jointly in day-ahead energy and AS markets.
    In \cite{Das2020}, optimal operation of wind-battery hybrid resources across day-ahead and real-time (balancing) energy markets was studied. 
    A robust model predictive control-based bidding strategy for wind-battery hybrid resources in real-time energy and AS markets was proposed in \cite{Xie2021}.
    These studies formulate multi-market participation as mathematical optimization problems, including Mixed Integer Linear Programming (MILP).
    However, these approaches typically rely on deterministic forecasts or explicitly specified uncertainty models for renewable generation and market prices, and their performance therefore depends on modeling assumptions. 
    As the complexity of multi-market participation increases, constructing appropriate uncertainty models becomes challenging.
    Alternatively, DRL-based methods have attracted increasing attention lately.  
    DRL enables the learning of operational policies through interaction with stochastic environments and directly optimizes expected long-term rewards~\cite{Powell2022}. 
    Moreover, such learning-based approaches can flexibly capture nonlinear system characteristics, such as battery degradation costs.
    Recent studies have therefore explored DRL-based methods for multi-market participation of battery energy storage systems~\cite{Dong2021,Anwar2022,Li2024:MultiMarket,Kortmann2025} and renewable-battery systems~\cite{Huang2021,Li2023:WindAncillaryDRL,Cardo-Miota2025}.
    
    \item \emph{Analyses of market participation models}: 
    A limitation of the above works is that they analyze the operation of renewable-battery systems under specific market settings and do not compare alternative market-participation structures. 
    While the term ``hybrid'' is often used loosely, we follow the definition in \cite{Ahlstrom2021}, in which a \emph{hybrid resource} participates in the market as a single integrated resource, whereas \emph{co-located resources} share the same interconnection point but are registered and bid as separate resources. 
    These two configurations are increasingly relevant in practice, and, for example,  California Independent System Operator (CAISO) supports both co-located and hybrid participation models~\cite{CAISO2025}, whereas Midcontinent Independent System Operator (MISO) allows only the co-located configuration~\cite{MISO2025}. 
    Although industry-facing studies~\cite{Kahrl2021,Ericson2022} qualitatively discuss the operational and market implications of these configurations, optimization-based analyses remain limited. 
    Existing quantitative studies, including the comparison of joint versus disjoint operation in \cite{Gomez2017} and the recent analysis of co-located versus hybrid participation models in \cite{Bhattacharjee2025}, mainly focus on energy-market settings and do not evaluate how different participation configurations influence market incentives, operational flexibility, or profitability across multiple revenue streams. 

    \item \emph{Design optimization of hybrid resources}: 
    Optimal sizing of renewable-battery systems has been extensively studied as reviewed in \cite{Agajie2023}. 
    Several techno-economic analysis tools have become widely used for plant-sizing studies, including commercial software such as HOMER Pro~\cite{HomerPro} and EnergyPRO~\cite{EnergyPro}, as well as the open-source Hybrid Optimization and Performance Platform (HOPP) developed by NREL~\cite{Guittet2022}.
    However, these studies and tools typically consider revenues only from energy markets or focus on standalone systems, and they do not explicitly model participation in multiple electricity markets when determining optimal capacities~\cite{Gupta2025}. 
    To the best of our knowledge, the recent work by Gupta \emph{et al.}~\cite{Gupta2025} is the only study that explicitly incorporates participation in multiple electricity markets when determining hybrid-resource capacities. 
    Their framework combines a MILP-based operational optimization with a global optimization algorithm to determine the optimal wind-battery configuration under the assumption of perfect forecasts of price and weather data. 
    While this approach provides an important step toward multi-market-aware hybrid resource sizing, the integration of uncertainty modeling remains unexplored. 
\end{enumerate}

\subsection{Contributions}
\label{sec:contributions}

Despite the growing literature summarized above, important gaps remain in both the operational optimization and design optimization of renewable-battery hybrid resources. 
This paper contributes to addressing the following three research gaps.

First, while DRL has become an attractive tool for addressing the increasing complexity of multi-market operational decisions under uncertainty, most existing studies focus primarily on battery energy storage systems~\cite{Dong2021,Anwar2022,Li2024:MultiMarket,Kortmann2025}. 
Although renewable-battery systems have been considered in~\cite{Huang2021,Li2023:WindAncillaryDRL,Cardo-Miota2025}, the operational constraints associated with AS provision remain insufficiently addressed.
In~\cite{Li2023:WindAncillaryDRL,Cardo-Miota2025},  such constraints are incorporated only as soft constraints through penalty terms in the reward function.
However, tuning such penalty weights so that constraints are satisfied without distorting the original reward maximization objective is notoriously difficult in DRL~\cite{Achiam2017,Ray2019}.
A promising approach has been proposed in~\cite{Huang2021}, where AS bidding is restricted through hard constraints on the action space of the DRL agent. 
Although this method provides a principled way to enforce operational feasibility, it considers AS provision only from battery storage and assumes perfect renewable generation forecasts.
This paper extends the framework of~\cite{Huang2021} to  develop a constraint-aware DRL framework that enables joint AS participation of renewable generation and battery storage under hard operational constraints. 
The proposed method also accounts for renewable forecast errors and captures the ability of battery storage to mitigate energy market imbalances, enabling a more realistic quantification of the operational and economic value of hybrid resources.

Second, although the distinction between hybrid and co-located resources has attracted increasing attention in recent years, quantitative analyses of their operational and economic differences remain limited.
Existing comparisons~\cite{Gomez2017,Bhattacharjee2025} primarily focus on energy market settings and do not consider participation across multiple revenue streams.
In this paper, we provide a systematic quantitative comparison between hybrid and co-located participation models under multi-market settings.
Using the proposed DRL-based operational framework, we evaluate how the two configurations differ in terms of operational flexibility, market participation strategies, and resulting economic performance.
This analysis provides new insights into how market-participation structures influence the value of renewable-battery hybrid resources.

Finally, as the central contribution of this paper, we address the design optimization of PV-battery hybrid resources while accounting for multi-market participation under uncertainty. 
The recent work by Gupta \emph{et al.}~\cite{Gupta2025} represents an important step toward multi-market-aware hybrid resource sizing by coupling MILP-based operational optimization with global search for capacity determination.
However, their approach relies on deterministic forecasts, and therefore does not explicitly account for uncertainty in renewable generation or market prices.
In parallel, recent studies have proposed DRL-based co-optimization frameworks for PV-battery systems~\cite{Cauz2024,Mantani2025:TEMPR}, in which a DRL agent simultaneously determines system capacities and operational policies.
Such approaches learn operational policies through interaction with stochastic environments while treating system capacities as part of the decision variables.
This allows design parameters to be optimized together with operational strategies, enabling direct optimization of long-term economic performance.
However, these studies primarily consider participation in energy markets and do not account for multiple revenue streams.
This paper extend the DRL-based co-optimization framework proposed in \cite{Mantani2025:TEMPR} to explicitly account for multi-market participation of PV-battery hybrid resources under uncertainty.

PV-battery systems are of particular interest because the design of these systems is strongly influenced by inverter capacity limits, which can lead to PV generation curtailment through inverter clipping.
DC-coupled battery storage can recover part of this otherwise curtailed energy, creating a strong coupling between PV capacity, battery sizing, and inverter utilization as discussed in~\secref{sec:problem_description}. 

\subsection{Organization}

The remainder of this paper is organized as follows. 
\Secref{sec:problem_description} introduces the hybrid resource design problem and relevant architectural characteristics. 
\Secref{sec:market_model} describes the market participation model considered in this study. 
\Secref{sec:algorithm} presents the proposed DRL-based co-optimization framework. 
\Secref{sec:simulation} provides case studies illustrating the effectiveness and implications of the proposed framework. 
Finally, \Secref{sec:conclusions} concludes the paper.


\section{Problem Description}
\label{sec:problem_description}

This section provides the conceptual and mathematical definitions of the hybrid resource design problem considered in this paper. 
We first introduce the background and terminology regarding hybrid and co-located resources in \secref{sec:definitions} and then discuss the key advantages of hybrid resources over conventional co-located configurations in \secref{sec:hybrid_benefits}. 
These insights are formulated into a design problem in \secref{sec:formulation}.


\subsection{Background and Definitions} 
\label{sec:definitions}

\begin{figure}
    \centering
    \begin{minipage}[t]{0.5\linewidth}
        \centering
        \includegraphics[width=\linewidth]{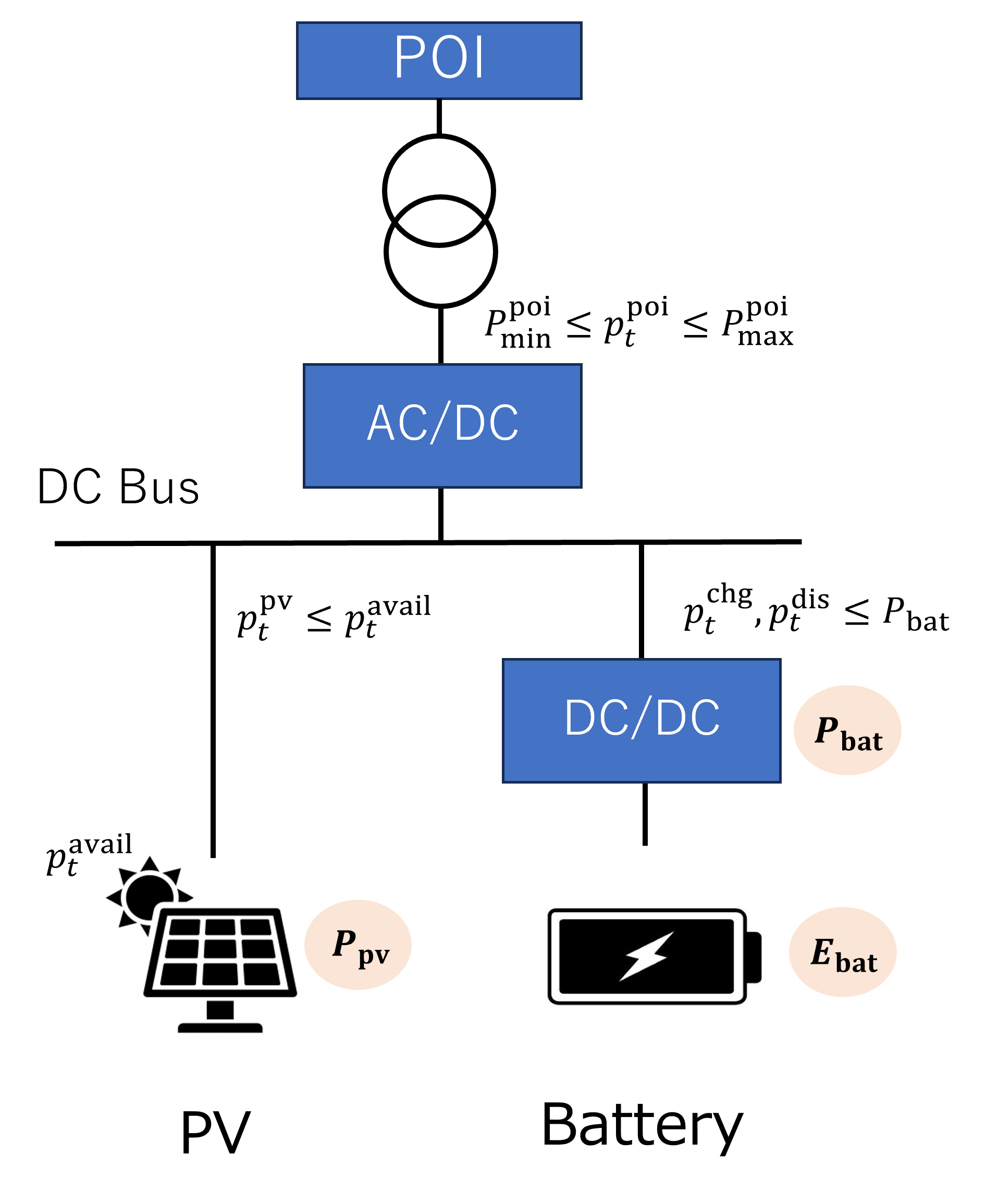}
        \subcaption{DC-coupled system}\label{fig:dc_coupled_colocated}
    \end{minipage}\hfill
    \begin{minipage}[t]{0.5\linewidth}
        \centering
        \includegraphics[width=\linewidth]{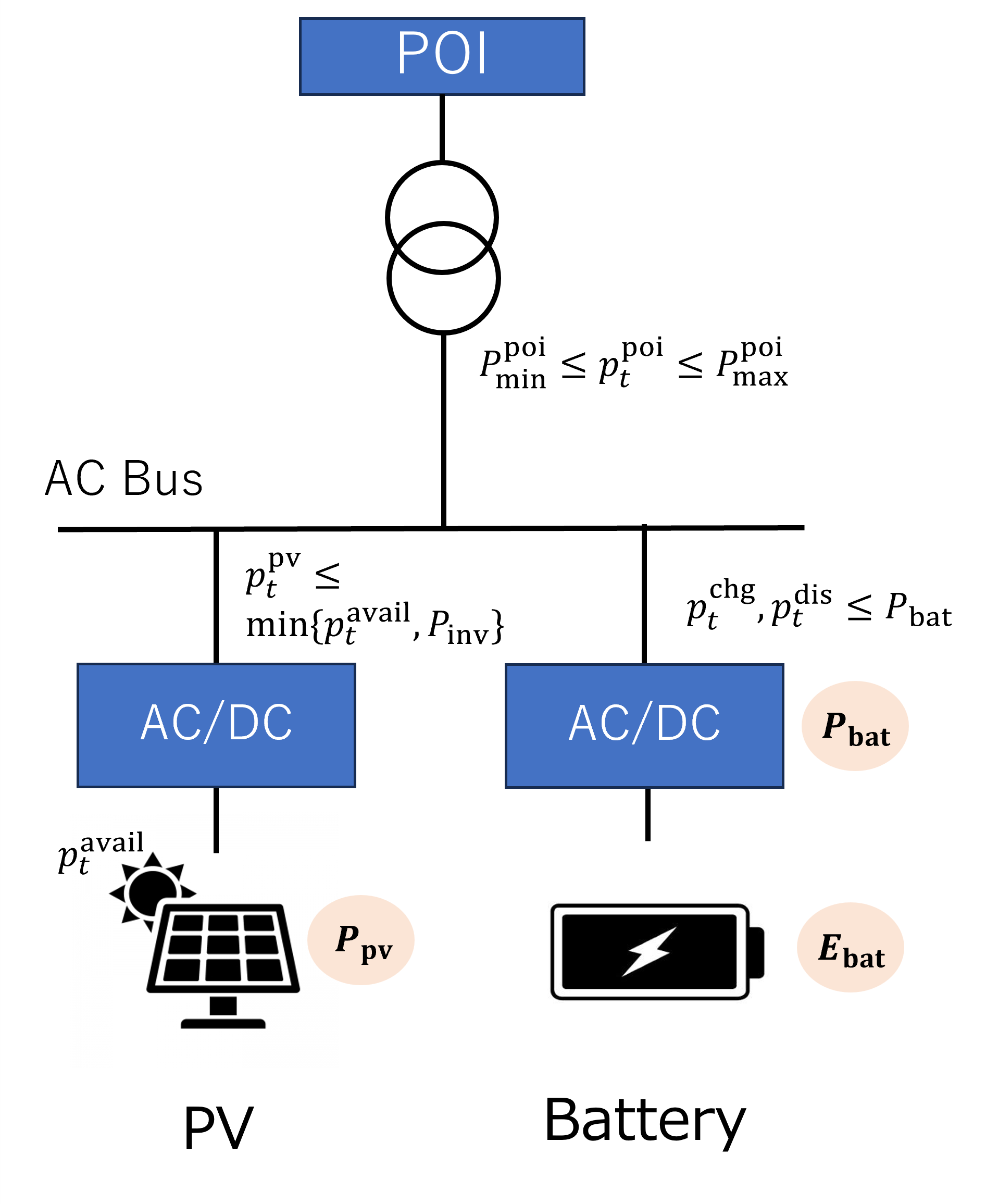}
        \subcaption{AC-coupled system}
        \label{fig:ac_coupled_colocated}
    \end{minipage}
    \caption{Comparison of PV-battery coupling architectures}
    \label{fig:ac_dc_coupling}
\end{figure}

PV-battery systems can be deployed under several architectural and market-participation configurations.
Despite market-specific differences, two key dimensions determine how these systems interact with the grid: 
\begin{itemize}[itemsep=0pt, topsep=2pt]
 \item Electrical configuration: How PV and battery are interconnected behind the Point of Interconnection (POI) to the grid.
 A DC-coupled configuration integrates PV and battery on the DC side, whereas an AC-coupled configuration uses separate inverters for PV and battery (see \figref{fig:ac_dc_coupling}).
 \item Market participation: How the PV-battery system is represented in the electricity market. 
  Co-located resources bid as separate PV and storage units, whereas hybrid resources bid and operate as a single dispatchable entity.
\end{itemize}
These two dimensions are conceptually independent, but historically they have been closely aligned in practice.
AC-coupled systems have long been the dominant implementation due to their architectural and regulatory simplicity, and have typically participated as co-located resources~\cite{Ahlstrom2021}.
In contrast, DC-coupled systems integrate PV and battery behind a shared inverter, making it impractical to meter PV and battery outputs separately for market purposes~\cite{CAISO2025}, and therefore requiring participation in the electricity market as a single entity.
Despite additional system and operational requirements associated with single-entity participation, DC-coupled hybrid resources are increasingly adopted due to their advantages as discussed in \secref{sec:hybrid_benefits}.
Motivated by this alignment, this study focuses on AC-coupled co-located resources and DC-coupled hybrid resources, hereafter referred to simply as \emph{co-located} and \emph{hybrid} resources, respectively.

\subsection{Advantages of Hybrid Resources} 
\label{sec:hybrid_benefits}

The DC-coupled architecture of hybrid resources provides several advantages over the AC-coupled architecture of co-located resources.
In particular, three key advantages identified in~\cite{Ahlstrom2021} are the recovery of clipped PV energy, improved output controllability, and enhanced reconfigurability.
These advantages are discussed below.

\begin{figure}
    \centering
    \includegraphics[width=0.9\linewidth]{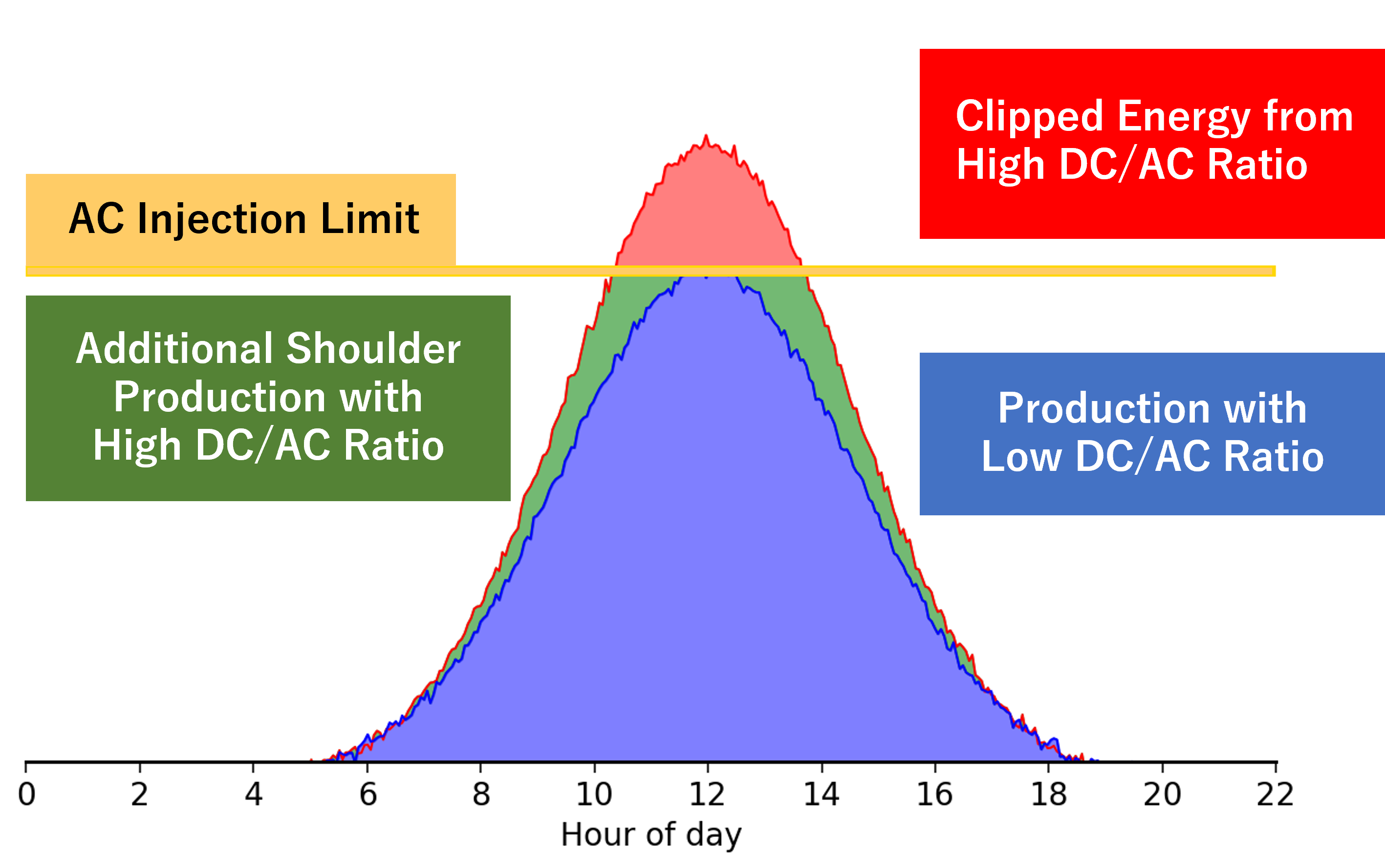}
    \caption{Recovery of clipped energy in hybrid resources}
    \label{fig:clipped_energy}
\end{figure}

First, hybrid architectures enable the recovery of ``clipped'' PV energy in plants with a high DC/AC ratio, as illustrated in \figref{fig:clipped_energy}.
Increasing the DC/AC ratio, \ie, oversizing the PV array relative to the inverter rating, expands PV production during shoulder hours, shown by the green region relative to the baseline low-ratio production shown by blue.
However, during peak irradiance periods, instantaneous PV output exceeds the inverter’s AC injection limit (yellow line), resulting in clipped energy in AC-coupled systems, indicated by the red region. 
Because an AC-coupled plant lacks a DC-side pathway between PV and storage, this excess energy must be curtailed.
In contrast, a DC-coupled configuration can route the clipped PV energy directly to the battery through the shared DC bus, recovering energy that would otherwise be lost.

Second, hybrid resources improve the controllability of net output by internally mitigating short-term fluctuations and forecast errors in PV production. Because the hybrid plant is dispatched as a single resource, the battery can offset real-time deviations between expected and realized PV output before they appear at the POI. This internal balancing reduces the magnitude of imbalance between market commitments and actual delivered power, thereby lowering exposure to imbalance penalties. 
The degree of mitigation depends on the available battery capacity, highlighting the need to consider battery sizing jointly with operational strategies.

Third, DC-coupled hybrid architectures offer a high degree of reconfigurability.
Because PV arrays and storage share a single grid-facing AC inverter, internal expansions or reconfigurations can often be implemented without altering the electrical characteristics seen at the POI.
This enables the resource to adapt to changes in market rules, regulatory requirements, or ancillary service definitions without triggering new interconnection studies, which are typically time-consuming and costly.
This advantage highlights the importance of optimal sizing studies for hybrid resources in adapting to market and regulatory changes.

Motivated by these three advantages, this paper formulates a design problem with a market-participation model that enables a quantitative and comparable evaluation of hybrid and co-located resources. 
Their implications are discussed through numerical experiments in \secref{sec:simulation}.

\subsection{Design Problem Formulation}
\label{sec:formulation}

Reflecting the reconfigurable nature of DC-coupled hybrid resources, we treat the grid-facing AC inverter as fixed by interconnection requirements and focus on optimizing the sizing of components on the DC side. 
Accordingly, the design vector $\omega$ is defined as
\begin{equation}
    \omega = \left\{ P_\mathrm{pv},\; E_\mathrm{bat},\; P_\mathrm{bat} \right\},
    \label{eq:design_variable}
\end{equation}
where $P_\mathrm{pv}$ stands for the nameplate capacity of the PV array, $E_\mathrm{bat}$ for the battery energy capacity, and $P_\mathrm{bat}$ for the power rating of the battery-side DC/DC converter as illustrated in \figref{fig:dc_coupled_colocated}.
With this configuration, the power injection at the POI at time $t$, denoted by $p_t^\mathrm{poi}$, is constrained by
\begin{equation}
    P_\mathrm{min}^\mathrm{poi} \le p_t^\mathrm{poi} \le P_\mathrm{max}^\mathrm{poi},  
    \label{eq:poi_limit}
\end{equation}
where the upper bound $P_\mathrm{max}^\mathrm{poi}>0$ is typically equal to the rated capacity of the grid-facing inverter, while the lower bound $P_\mathrm{min}^\mathrm{poi}$ depends on whether the inverter is unidirectional ($P_\mathrm{min}^\mathrm{poi}=0$) or bidirectional ($P_\mathrm{min}^\mathrm{poi}=-P_\mathrm{max}^\mathrm{poi}$).
The internal DC-side power balance is given by
\begin{equation}
    p_t^\mathrm{poi} = p_t^\mathrm{pv} - p_t^\mathrm{chg} + p_t^\mathrm{dis},
    \label{eq:dc_power_balance}
\end{equation}
where $p_t^\mathrm{pv}$ stands for the PV output power at time $t$, bounded by the available PV generation $p_t^\mathrm{avail}$ determined by solar irradiance, i.e., 
\begin{align}
 0 \le p_t^\mathrm{pv} \le p_t^\mathrm{avail}, 
\end{align}
and $p_t^\mathrm{chg}$ and $p_t^{\mathrm{dis}}$ represent the battery charging and discharging powers, respectively.
The battery-side converter constrains the charging and discharging powers as
\begin{align}
    0 \le p_t^{\mathrm{chg}} \le P_\mathrm{bat}, \label{eq:converter_limit_chg} \\
    0 \le p_t^{\mathrm{dis}} \le P_\mathrm{bat}. \label{eq:converter_limit_dis}
\end{align}
The battery State-of-Charge (SOC), denoted by $x_t^\mathrm{soc}$, evolves according to
\begin{equation}
    x^\mathrm{soc}_{t+1}
    = x^\mathrm{soc}_t
      + \eta_\mathrm{c} \frac{p_t^\mathrm{chg}\Delta t}{E_\mathrm{bat}}
      - \frac{1}{\eta_\mathrm{d}}\frac{p_t^\mathrm{dis}\Delta t}{E_\mathrm{bat}},
    \label{eq:soc_dynamics}
\end{equation}
subject to the SOC bounds
\begin{equation}
    S_{\min} \le x^\mathrm{soc}_t \le S_{\max}, \label{eq:soc_limit}
\end{equation}
where $\eta_\mathrm{c}$ and $\eta_\mathrm{d}$ represent charging and discharging efficiencies, respectively, $\Delta t$ is the simulation time step, and $S_{\min}$ and $S_{\max}$ specify the allowable SOC limits.

The economic performance of the hybrid resource is determined by the total capital and operational expenditures and the total market revenue over the considered evaluation period. 
In this paper, the design problem is formulated for a representative target year to focus on the structural differences between hybrid and co-located resources with a simple problem setting. 
The proposed formulation can be extended in a straightforward manner to a multi-year evaluation framework based on Net Present Value (NPV), as in~\cite{Gupta2025}. 
By letting $F_\mathrm{mkt}^\mathrm{opt}(\omega)$ represent the expected annual revenue obtained under optimal market-participation behavior for a given design $\omega$,  
the design optimization problem is given by 
\begin{equation}
    \max_{\omega} 
    \;\; 
    F_\mathrm{mkt}^\mathrm{opt}(\omega)
    - C_{\mathrm{cap}}(\omega),
    \label{eq:design_opt}
\end{equation}
where $C_{\mathrm{cap}}(\omega)$ represents the annualized capital expenditure of the PV array, battery system, and inverter.
The joint optimization of design variables and market-participation decisions is detailed in \secref{sec:algorithm}.

\section{Market Model}
\label{sec:market_model}

This section presents the market model used in this paper. 
The purpose is not to reproduce the detailed rules of any specific market (\eg, CAISO, MISO, among others), but rather to capture the essential structure of multi-market participation in a simplified yet general form, while explicitly accounting for PV prediction errors.
To this end, we impose the following modeling assumptions: 
\begin{enumerate}[label=(A\arabic*), itemsep=0pt, topsep=2pt]
\item \label{ass:price_taker} \emph{Price-taker assumption}:  
The market participant is assumed to act as a price taker.
All market prices (energy and AS) are treated as exogenous signals and are not affected by the  submitted bids.
\item \label{ass:quantity_clearing}
\emph{Quantity-based bidding}:  
The model focuses on determining the optimal bid quantities for each market. Bid prices are not explicitly modeled and are assumed to be set in such a way that the submitted quantities are fully cleared. This abstraction allows us to concentrate on quantity allocation across markets without loss of generality for price-taker resources.
\item \label{ass:forward_energy}
\emph{Forward energy market representation}:  
Energy market participation is modeled as a forward energy commitment cleared at the same time scale as the procurement of AS. Deviations between the committed energy schedule and realized injection at the POI, arising from PV forecast errors and  operational constraints, result in energy imbalances that are penalized.
\item \label{ass:imbalance}
\emph{Treatment of energy imbalances}:  
Although energy imbalances may be resolved through closer-to-real-time energy trading in actual market settings, they are aggregated and represented as ex-post imbalance settlements with penalty costs.
\item \label{ass:ancillary}
\emph{AS commitments}:  
AS bids are represented as capacity availability commitments over the corresponding delivery interval, and performance-based remuneration associated with real-time activation is not explicitly modeled.
The committed capacity is required to remain available throughout the interval, thereby imposing operational constraints on the hybrid resource.
\item \label{ass:capacity_payment}
\emph{Capacity remuneration}:  
Capacity payments represent remuneration for firm capacity provision. Eligibility for capacity remuneration is assumed to be determined solely by the installed PV and storage capacities, independent of operational outcomes. While market rules may impose must-offer or availability obligations on capacity-qualified resources, such obligations are assumed to be satisfied by design and are not modeled. 
\end{enumerate}
Assumptions~\ref{ass:price_taker} and~\ref{ass:quantity_clearing} are standard in resource design studies.
The remaining assumptions are discussed later in this section.
Under these assumptions, the total market revenue of the hybrid resource can be decomposed as
\begin{equation}
    F_{\mathrm{mkt}}
    =
    W_\mathrm{anu}\sum_{t=1}^{T} \left( F_t^\mathrm{e}
    + F_t^\mathrm{as} \right)
    + F_\mathrm{cap},
    \label{eq:total_market_revenue}
\end{equation}
where $t=1,\dots,T$ indexes the market settlement intervals, $W_\mathrm{anu}$ represents the annualization factor, $F_t^\mathrm{e}$ and $F_t^\mathrm{as}$ represent the revenue from the energy and AS markets at time $t$, respectively, and $F_\mathrm{cap}$ represents capacity remuneration. 
Details of each term are provided below.

\subsection{Energy Market} \label{sec:energy_market}

The hybrid resource submits an energy bid $b_t^{\mathrm{e}}$ for each time slot $t$, representing the scheduled net power injection at the POI during interval $t$. 
Depending on the market design, $b_t^\mathrm{e}$ may correspond to a day-ahead schedule, an hour-ahead schedule, or another form of forward energy commitment, and the clearing is temporally aligned with the gate closure of AS markets under Assumption~\ref{ass:forward_energy}.
Given the stochastic nature of renewable generation, deviations between the scheduled energy bid $b_t^\mathrm{e}$ and the actual power available at the POI may arise. 
These deviations are commonly referred to as energy imbalances, and may be resolved through closer-to-real-time energy trading or settled according to market-specific imbalance settlement rules.
A common accounting representation of market revenue decomposes payments into a forward settlement and an imbalance settlement as 
\begin{align}
    F_t^{\mathrm{e}}
    &=
    \lambda_t^\mathrm{e} b_t^\mathrm{e} \Delta t
    +
    \lambda_t^\mathrm{imb} (x_t^\mathrm{e} - b_t^\mathrm{e}\Delta t),
    \label{eq:DA_RT_revenue}
\end{align}
where $x_t^\mathrm{e}$ represents the realized net energy injection over interval $t$, 
and $\lambda_t^\mathrm{e}$ and $\lambda_t^\mathrm{imb}$ are the market clearing price and the settlement price applied to the energy imbalance, respectively.
In the literature, two conceptually distinct approaches can be found for modeling the economic impact of the imbalances.
One class of studies models real-time energy prices explicitly, and evaluates deviations through the resulting price differences between forward and real-time markets (\eg, \cite{Rahimiyan2016,Mehdipourpicha2021}).  
Another class of studies instead treats deviations themselves as an economic risk, modeling the effect of price differences through an imbalance penalty applied to the mismatch between scheduled and realized injections (\eg, \cite{Jeong2023,Yang2023}). 
In the former modeling paradigm, deviations between forward and real-time prices may, in some market conditions, lead to arbitrage opportunities~\cite{Mehdipourpicha2021}.
However, a systematic investigation of such inter-market arbitrage effects is beyond the scope of the present study and is left for future work.
Instead, under assumption \ref{ass:imbalance}, we use  the latter imbalance-penalty-based formulation, which provides a simple representation of the economic consequences of energy deviations: 
\begin{equation}
F_t^\mathrm{e}
= \lambda_t^\mathrm{e}
\left(
x_t^\mathrm{e}
- \pi_t^{\mathrm{imb}}
\left| \Delta E_t \right|
\right),
\label{eq:energy_penalty_model}
\end{equation}
with the energy imbalance $\Delta E_t$ defined by 
\begin{align}
    \Delta E_t= x_t^\mathrm{e} - b_t^\mathrm{e}\Delta t,
\end{align}
where $\pi_t^{\mathrm{imb}} \ge 0$ is penalty factor. 
The formulation in \eref{eq:energy_penalty_model} is equivalent to \eref{eq:DA_RT_revenue} when the $\lambda_t^\mathrm{imb}$ is chosen such that 
\begin{align}
    \lambda_t^\mathrm{imb} = 
        \begin{cases}
        \lambda_t^\mathrm{e} (1 - \pi_t^{\mathrm{imb}}), & \Delta E_t > 0, \\ 
        \lambda_t^\mathrm{e} (1 + \pi_t^\mathrm{imb}), & \Delta E_t \le 0. 
    \end{cases}
\end{align}

\subsection{Ancillary Service Markets} \label{sec:as_markets}

In addition to the energy market, various AS markets provide services to support system frequency and voltage stability.
Among them, this paper focuses on two representative classes of AS: contingency reserves and regulation services. 
These services are directly associated with system frequency control and are commonly procured across most electricity markets, although their design and settlement rules vary across system operators and regions~\cite{NERC2021}.
Contingency reserves are intended to respond to infrequent but severe system disturbances, such as the sudden loss of a large generator or an interconnection with a neighboring system.
They are activated rarely but require sufficient deliverable power and energy to be available over a prescribed activation duration.
In contrast, regulation services are designed to continuously compensate for small power imbalances arising from load variations and forecast errors.
They are typically implemented through automatic or semi-automatic control mechanisms, such as Automatic Generation Control (AGC)~\cite{Brooks2019}, and involve frequent adjustments in both upward and downward directions to maintain system frequency within acceptable limits.

Following common approaches in the literature on AS market modeling \cite{Dong2021,Anwar2022,Li2024:MultiMarket}, this paper uses a capacity-based formulation.
In particular, we model the AS revenue based on three types of capacity bids of contingency reserve $b_t^\mathrm{res}$, upward regulation $b_t^\mathrm{up}$, and downward regulation $b_t^\mathrm{dn}$: 
\begin{align}
\label{eq:AS_market}
F_t^\mathrm{as}
=
\lambda_t^\mathrm{res} b_t^\mathrm{res}
+
\lambda_t^\mathrm{up} b_t^\mathrm{up}
+
\lambda_t^\mathrm{dn} b_t^\mathrm{dn},
\end{align}
where $\lambda_t^\mathrm{res}$, $\lambda_t^\mathrm{up}$, and $\lambda_t^\mathrm{dn}$ represent the capacity prices for contingency reserves, upward regulation, and downward regulation services, respectively.
While practical AS markets often include multiple subcategories depending on response time and minimum duration, such distinctions can be handled within the same modeling framework considered here. 
In addition to capacity-based remuneration, AS markets may offer various ex-post settlement mechanisms. 
For regulation services in particular, remuneration may depend on realized performance metrics, and make-whole payments may be used to compensate for opportunity costs relative to energy market participation~\cite{Brooks2019}.
However, under Assumption~\ref{ass:ancillary}, this paper does not explicitly model such ex-post remuneration adjustments and  focuses on the ex-ante capacity commitment, as represented in~\eref{eq:AS_market}.
Committing AS capacity 
 implies that sufficient energy 
must be reserved to ensure deliverability over a prescribed activation duration. 
While the actual AS activations, including whether a contingency event occurs or the magnitude and direction of AGC signals, are unknown at the time of market bidding, this paper assumes, following~\cite{He2016,Dong2021}, that prescribed duration parameters for contingency reserves ($H_{\mathrm{res}}$), upward regulation ($H_{\mathrm{up}}$), and downward regulation ($H_{\mathrm{dn}}$) are specified a priori, representing worst-case durations for sustaining the committed capacity without violating energy constraints under any admissible activation profile.
The resulting power and energy feasibility constraints are explicitly incorporated into the system model in \secref{sec:rl_bidding}.

\subsection{Capacity Remuneration}\label{sec:capacity_market}

Capacity remuneration is determined based on an accredited firm capacity. 
In principle, the capacity contribution of a hybrid resource depends on its operational strategy, particularly on how battery is scheduled in conjunction with variable renewable generation~\cite{Ericson2022}.
However, market-wide and operation-aware capacity accreditation methods for hybrid resources are not yet fully established, and current capacity market frameworks typically rely on simplified, technology-specific eligibility rules without considering operational details (see, \eg,~\cite{CPUC2020}).
Therefore, under Assumption~\ref{ass:capacity_payment}, this paper models capacity remuneration based on the sum of the eligible capacities of individual components, rather than on their joint operational behavior: 
\begin{align}
F_{\mathrm{cap}}
=
\lambda^{\mathrm{cap}}
\min\left\{ P_\mathrm{max}^\mathrm{poi}, \, 
 P_\mathrm{cap}
\right\}
\label{eq:capacity_revenue}
\end{align}
where $\lambda^\mathrm{cap}$ represents the capacity remuneration price, and $P_\mathrm{cap}$ is given by
\begin{align}
 P_\mathrm{cap} = 
 \phi_\mathrm{pv} \, P_\mathrm{pv}
+
\min\!\left\{
 P_\mathrm{bat},
\frac{ E_{\mathrm{bat}}}{H_{\mathrm{cr}}}
\right\}. 
\end{align}
The first term represents the accredited capacity of PV generation.
Since PV output may not coincide with periods of system peak demand, its contribution to firm capacity is generally discounted relative to the installed capacity, by the factor $\phi^{\mathrm{pv}} \in [0,1]$ defined as the fraction of installed PV capacity that can be counted as firm capacity~\cite{Dent2016}. 
The second term represents the capacity of the battery storage.
The eligible capacity of battery is constrained by a minimum discharge duration requirement. 
This duration requirement is set to $H_\mathrm{cr}=\SI{4}{h}$ and is commonly referred to as the four-hour rule, which is adopted in markets such as CAISO.
Recent policy discussions have highlighted the limitations of short-duration storage, motivating proposals for longer duration requirements~\cite{Denholm2023}.

\section{Co-Optimization Methodology} \label{sec:algorithm}

This section presents the proposed method for the joint optimization of design decisions and market-participation behaviors based on DRL. 
The overall optimization framework is first outlined in \secref{sec:co-optimization}, and the detailed descriptions used to learn market-participation policies and designs are provided in \secref{sec:rl_bidding} and \secref{sec:design_update}, respectively.

\subsection{Overall Co-Optimization Framework}
\label{sec:co-optimization}

The co-optimization method adopted in this paper is based on our previous work~\cite{Mantani2025:TEMPR} and is extended to evaluate the economic performance of hybrid resources participating in multiple electricity markets. 
In a standard DRL-based framework, system design parameters are fixed a priori, and the operational strategy is optimized under the given design. 
When design optimization is considered, the operational problem typically needs to be solved repeatedly for each candidate, resulting in a bi-level structure~\cite{Perera2019ML,Li2022,Kang2023,Pan2024}, where the upper-level explores the design space and the lower-level RL solves the operational decision problem. 
Although conceptually straightforward, this separation often leads to substantial computational burden~\cite{Mantani2025:TEMPR}. 
In contrast, the proposed framework integrates system design directly into the DRL training process.

\begin{figure}
    \centering
    \includegraphics[width=\linewidth]{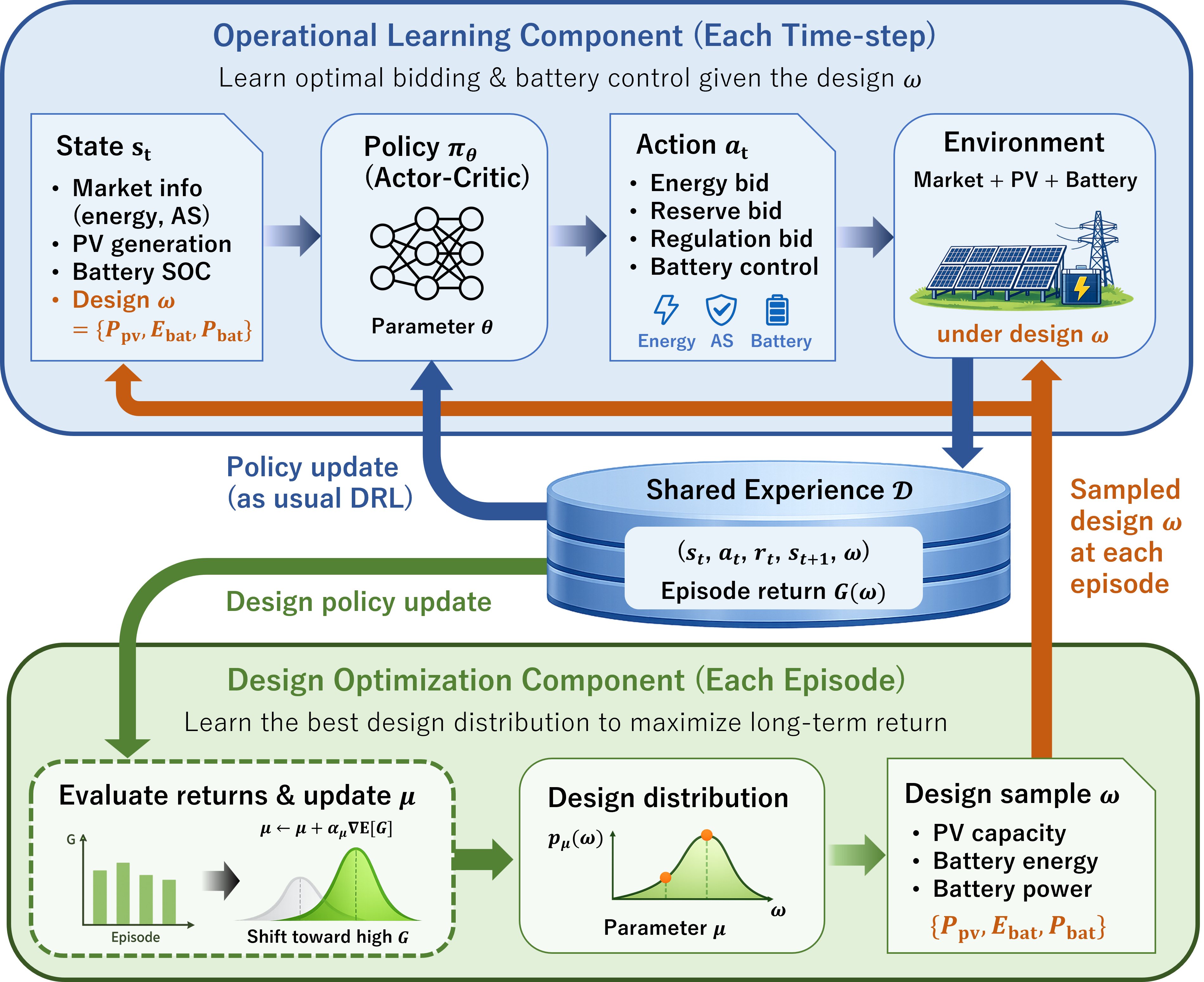}
    \caption{The overall co-optimization framework}
    \label{fig:framework}
\end{figure}

\Figref{fig:framework} illustrates the overall co-optimization framework. 
Let $\omega$ represent the system design parameter as in \eref{eq:design_variable}, and $\pi_\theta$ the operational policy parameterized by $\theta$. 
The upper part of the figure shows the operational learning component, which follows a standard DRL framework except that the design parameter $\omega$ is included in the state and remains fixed within each episode.
Given a sampled design $\omega$, the agent interacts with the environment, collects transition data, and stores them in the shared experience buffer $\mathcal{D}$. 
The policy parameter $\theta$ is then updated using mini-batches sampled from $\mathcal{D}$.
The lower part of the figure shows the design optimization component. 
At the beginning of each episode, a design instance $\omega$ is sampled from the distribution $p_\mu(\omega)$ parameterized by $\mu$. 
After executing the episode, the return $G(\omega)$ is evaluated and associated with the sampled design. 
Using the accumulated experience in $\mathcal{D}$, the distribution parameter $\mu$ is updated so as to increase the expected return $\mathbb{E}_{\omega \sim p_\mu}[G(\omega)]$, effectively shifting the design distribution toward regions yielding higher performance.
Through this structure, the policy parameter $\theta$ and the distribution parameter $\mu$ are updated from shared experience while operating at different time scales, enabling joint optimization of system design and operation within a single learning process.

\subsection{Operational Learning Component}
\label{sec:rl_bidding}

This subsection describes the operational learning component for multi-market participation. 
The observation, action, and reward spaces are defined in \secref{sec:state_space}, followed by the constraint handling strategy in \secref{sec:AS_provision}, and the modeling of real-time battery control and energy imbalances in \secref{sec:imbalances}.

\subsubsection{Observation, Action, and Reward} \label{sec:state_space}

In DRL, the interaction between an agent and an environment is typically modeled as a Markov Decision Process (MDP). 
At each time step $t$, the agent observes the system state $s_t$, selects an action $a_t$, receives a reward $r_t$, and the environment transitions to the next state $s_{t+1}$. 
The objective is to learn a policy that maximizes the expected cumulative reward.
In the market participation, however, key variables such as renewable generation and market prices are not fully observable in advance, making the problem a Partially Observable MDP (POMDP). 
In this paper, it is assumed that market and renewable variables up to time $t-1$ are observable when making a decision at time $t$. 
This setting corresponds, for example, to participation in hour-ahead markets ($\Delta t = 1$~h), where bids are determined based on the most recent observations\footnote{
While this assumption simplifies the exposition, the proposed formulation can be extended to settings with longer look-ahead requirements, such as day-ahead markets, by incorporating forecast information available at earlier decision stages.}. 
Accordingly, the agent receives an observation
\begin{equation}
    o_t = \left(p_{t-1}^\mathrm{avail}, \lambda_{t-1}^\mathrm{e}, \lambda_{t-1}^\mathrm{res}, \lambda_{t-1}^\mathrm{up}, \lambda_{t-1}^\mathrm{dn}, x^\mathrm{soc}_t; \omega \right),
\end{equation}
and the underlying state can be represented as a sequence of past observations:
\begin{equation}
    s_t = \left(o_0, o_1, \dots, o_t \right).
    \label{eq:state_representation}
\end{equation}

To address this partial observability, this paper introduces a Long Short-Term Memory (LSTM)~\cite{Zhou2019:LSTM_Gen,Zhou2019:LSTM_Price} to extract temporal features from past observations. 
Specifically, the time series of available PV generation and market prices
$(p^\mathrm{avail}_{t-1}, \lambda^\mathrm{e}_{t-1}, \lambda^\mathrm{res}_{t-1}, \lambda^\mathrm{up}_{t-1}, \lambda^\mathrm{dn}_{t-1})$
are processed by the LSTM. 
The resulting hidden state $h_t$ and cell state $c_t$ provide a compact representation of the observation history, allowing the state to be approximated as
\begin{equation}
    s_t \approx \left(h_t, c_t, x^\mathrm{soc}_t; \omega \right).
\end{equation}
A comparative study of several LSTM-based DRL agents for market participation problems was conducted in~\cite{Mantani2025:TEMPR}, where four candidate architectures were evaluated. 
Among them, the combination of Deep Deterministic Policy Gradient (DDPG)~\cite{Lillicrap2015:DDPG} with LSTM achieved the best performance in terms of cumulative reward. 
Therefore, this architecture is adopted in this study. 
Note that the proposed formulation is not restricted to DDPG and can be combined with other recurrent DRL algorithms.

The agent’s action specifies the desired bidding and battery operational decisions for market participation and is given by 
\begin{align}
  a_t = \left( 
     a_t^\mathrm{e}, a_t^\mathrm{res}, a_t^\mathrm{up}, a_t^\mathrm{dn}, a_t^\mathrm{imb} 
    \right),  
\end{align}
where all action components are defined as normalized variables taking values in $[0,1]$.
The components $a_t^\mathrm{e}$, $a_t^\mathrm{res}$, $a_t^\mathrm{up}$, and $a_t^\mathrm{dn}$ represent the agent’s intended fractions of the available capacity allocated sequentially to the energy market, contingency reserve, regulation-up service, and regulation-down service, respectively.
These fractions are converted into feasible bidding quantities by explicitly accounting for capacity and energy constraints, as detailed in \secref{sec:AS_provision}.
The variable 
$a_t^\mathrm{imb}$ governs how the available battery energy is allocated between mitigating real-time energy imbalances and preserving energy capacity for future arbitrage opportunities as described in \secref{sec:imbalances}. 

The operational objective is to maximize the total net revenue over a given time horizon $T$.
The reward $r_t$ obtained by the agent at each time step consists of revenues from the energy and AS markets minus the battery degradation cost:
\begin{align}
 r_t = F_t^\mathrm{e}
+ F_t^\mathrm{as}
- C_t^\mathrm{deg}, 
\end{align}
where $F_t^\mathrm{e}$ and $F_t^\mathrm{as}$ are defined in \eref{eq:energy_penalty_model}, and \eqref{eq:AS_market}, respectively, and $C_t^\mathrm{deg}$ represents the battery degradation cost.
The degradation in general depends nonlinearly on the SOC and cycling conditions~\cite{Han2014:Degradation,Kim2017:Degradation}.
For simplicity, however, this paper adopts a linearized formulation following~\cite{Jeong2023}:
\begin{align}
  C_t^\mathrm{deg} = \beta_\mathrm{deg} (p^\mathrm{chg}_t + p^\mathrm{dis}_t) \Delta t,
\end{align}
where $\beta_\mathrm{deg}$ is a constant.
Note that the proposed DRL-based framework is not restricted to this simplified model and can accommodate more detailed nonlinear degradation models without modifying the overall formulation.


\subsubsection{Constraint Handling Strategy}
\label{sec:AS_provision}

\begin{figure}
    \centering
    \includegraphics[width=0.95\linewidth]{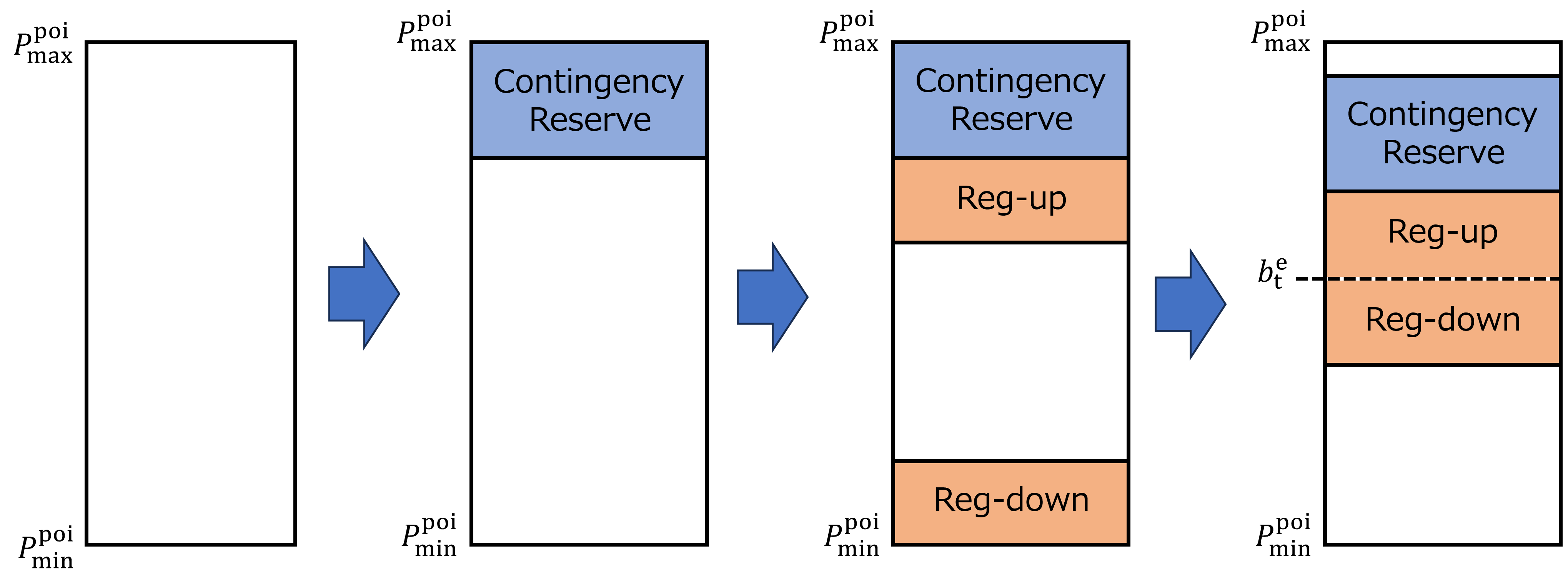}
    \caption{Schematic overview of the serial strategy}
    \label{fig:serial_strategy}
\end{figure}

To ensure compliance with AS requirements under renewable forecast uncertainty and operational constraints, the agent actions are transformed into feasible AS bidding quantities. 
A serial capacity allocation strategy proposed in~\cite{Huang2021} is adopted for this purpose. 
As illustrated in \figref{fig:serial_strategy}, AS capacities are allocated sequentially so that the capacity and energy margins reserved for earlier services are explicitly accounted for when determining the feasible bids for subsequent services. 
After the feasible AS capacities are determined, the remaining feasible operating range is used to determine the energy bid $b_t^\mathrm{e}$, which represents the scheduled net power injection at the POI. 
In this paper, this strategy is extended from battery-only AS provision in~\cite{Huang2021} to PV-battery hybrid resources, where both PV and battery can contribute to AS provision under PV forecast uncertainty.


Among three AS bids, the capacity for contingency reserve is first allocated.
For simplicity of exposition, we assume that the contingency reserve is provided exclusively by the battery, without loss of generality\footnote{From a modeling perspective, PV-based reserve provision can be incorporated using the same framework as that for regulation-up services introduced later. Moreover, in practice, contingency events are infrequent, and allocating PV capacity would require preventive curtailment, leading to unnecessary economic losses.}.
In this case, the feasible reserve capacity is constrained by the POI capacity in \eref{eq:poi_limit} and the discharge power limit of the battery converter in \eref{eq:converter_limit_dis}, as well as by the dischargeable energy available under the SOC limit in \eref{eq:soc_limit}.
By using the agent action $a_t^{\mathrm{res}}$ specifying the desired fraction committed to reserve, the  contingency reserve bid $b_t^{\mathrm{res}}$ is given by 
\begin{align}
  b_t^{\mathrm{res}} 
  = \min\left\{ 
    a_t^{\mathrm{res}} P_\mathrm{poi}, \, 
    P_\mathrm{bat}, \, 
    \frac{ E_t^{\mathrm{bat}\uparrow} }{ H_\mathrm{res} } 
  \right\}, 
\end{align}
where $P_\mathrm{poi} = P_\mathrm{max}^\mathrm{poi} - P_\mathrm{min}^\mathrm{poi}$ represents the total POI capacity available for allocation, $E_t^{\mathrm{bat}\uparrow}$ represents the dischargeable energy available in the battery given by
\begin{align}
  E_t^\mathrm{bat\uparrow} & = 
  \eta_\mathrm{d} E_\mathrm{bat}\bigl(x_t^{\mathrm{soc}} - S_{\min}\bigr),
\end{align}
and $H_\mathrm{res}$ is the required reserve duration specified in \secref{sec:as_markets}.

After allocating the contingency reserve capacity, we next determine the regulation-up bid by accounting for the remaining capacity and energy margins. 
The contribution of the battery to regulation-up provision is limited by the remaining converter power capacity and dischargeable energy. 
Thus, the maximum upward capacity $M_t^\mathrm{bat\uparrow}$ from the battery is given by 
\begin{align}
    M_t^\mathrm{bat\uparrow}  = \min\left\{ P_\mathrm{bat}-b_t^\mathrm{res}, \, \frac{ E_t^{\mathrm{bat}\uparrow} -b_t^\mathrm{res}H_\mathrm{res}}{H_\mathrm{up}}\right\}. 
    \label{eq:battery_dis_margin}
\end{align}
For PV generation, the available capacity is limited by forecast uncertainty.
Let $P_t^\mathrm{pred}$ represents the predicted available PV power obtained from the LSTM-based predictor.
We introduce a reliability parameter $\kappa \in [0,1]$ that represents the fraction of the predicted PV power that is conservatively assumed to be available for AS provision.
Thus, the guaranteed PV capacity is given by 
\begin{align}
  M_t^\mathrm{pv} &= \kappa P_t^\mathrm{pred}.  
  \label{eq:pv_for_AS}
\end{align}
As a result, by using the agent action $a_t^{\mathrm{up}}$, which specifies the desired fraction of the remaining capacity, the feasible regulation-up bid $b_t^\mathrm{up}$ is given by
\begin{align}
   b_t^\mathrm{up} = \min\left\{ 
     a_t^{\mathrm{up}}\left(P_\mathrm{poi}-b_t^\mathrm{res}\right), \, 
     M_t^\mathrm{bat\uparrow} + M_t^\mathrm{pv}
  \right\}.
\end{align}
Here, the first term enforces the remaining POI capacity limit, while the second term represents the total regulation-up capacity available from the battery and PV generation.

Finally, the regulation-down bid $b_t^\mathrm{dn}$ is determined from the remaining capacity and downward energy margins.
The downward regulation capability of the battery is constrained by the charging power limit of the converter in \eref{eq:converter_limit_chg} and the chargeable energy available under the SOC limit in \eref{eq:soc_limit}. 
Thus, the maximum downward capacity $M_t^\mathrm{bat\downarrow}$ from the battery is given by 
\begin{align}
    M_t^\mathrm{bat\downarrow} = \min\left\{ P_\mathrm{bat}, \, \frac{ E_t^{\mathrm{bat}\downarrow}}{H_\mathrm{dn}}\right\}. 
\end{align}
where $E_t^{\mathrm{bat}\downarrow}$ represents the chargeable energy given by
\begin{align}
   E_t^\mathrm{bat\downarrow} & =  E_\mathrm{bat}\bigl(S_{\max} - x_t^{\mathrm{soc}}\bigr)/\eta_\mathrm{c}.
\end{align}
In addition, the PV generation that is not reserved for regulation-up service can contribute to regulation-down provision.
Specifically, when the regulation-up bid exceeds the maximum capacity that can be supplied by the battery, \ie, $b_t^\mathrm{up} > M_t^\mathrm{bat\uparrow}$, the excess portion of the regulation-up commitment must be provided by PV generation.
Accordingly, the amount of PV capacity reserved for regulation-up service is given by
\begin{align}
  b_t^\mathrm{up,pv} = \max\left\{ b_t^\mathrm{up} - M_t^\mathrm{bat\uparrow}, 0 \right\}. 
  \label{eq:pv_up_regulation}
\end{align}
Then, by using the agent action $a_t^\mathrm{dn}$, the feasible regulation-down bid $b_t^\mathrm{dn}$ is given by 
\begin{align}
 b_t^{\mathrm{dn}} = 
 \min\Bigl\{ 
 & a_t^{\mathrm{dn}}\left(P_\mathrm{poi} - b_t^{\mathrm{res}} - b_t^{\mathrm{up}}\right), \,  
 \notag \\
 & 
 \quad M_t^\mathrm{bat\downarrow} +  M_t^\mathrm{pv} - b_t^\mathrm{up,pv} 
 \Biggr\}.
\end{align}

In addition to AS markets, the hybrid resource submits its bid to the forward energy market. 
The energy market bid $b_t^\mathrm{e}$ corresponds to the scheduled power injection at the POI, and is subject to the POI power limits considering the AS commitments as 
\begin{align}
 P_\mathrm{min}^\mathrm{poi} +b_t^\mathrm{dn} 
 \le b_t^\mathrm{e} \le 
 P_\mathrm{max}^\mathrm{poi} - b_t^\mathrm{res} -b_t^\mathrm{up}. 
 \label{eq:energy_POI_limit}
\end{align}
In addition, to reduce energy imbalances, it is desirable to limit the energy market bid based on the PV prediction $P_t^\mathrm{pred}$ and the available battery margins.
Specifically, we restrict the feasible bidding range as
\begin{align}
P_t^\mathrm{pred} - b_t^\mathrm{up,pv} - B_t^{\mathrm{chg}}
\le b_t^\mathrm{e}
\le
P_t^\mathrm{pred} - b_t^\mathrm{up,pv} + B_t^{\mathrm{dis}}
\label{eq:energy_feasible}
\end{align}
where $B_t^\mathrm{dis}$ and $B_t^\mathrm{chg}$ represent the battery chargeable and dischargeable  margins after allocating the AS bids, respectively, and given by 
\begin{align}
  B_t^\mathrm{dis} = \min\bigl\{ 
     & P_\mathrm{bat} -b_t^\mathrm{res} - b_t^\mathrm{up,bat},\, \notag \\ 
     & \bigl( E_t^\mathrm{bat\uparrow} - b_t^\mathrm{res}H_\mathrm{res} -b_t^\mathrm{up,bat}H_\mathrm{up} \bigr)/\Delta t \bigr\},  
  \\
  B_t^\mathrm{chg} = \min\bigl\{ 
     & P_\mathrm{bat} - b_t^\mathrm{dn,bat},\, \notag \\ 
     & \bigl( E_t^\mathrm{bat\downarrow} -b_t^\mathrm{dn,bat}H_\mathrm{dn} \bigr)/\Delta t \bigr\}.  \label{eq:B_chg}
\end{align}
where $b_t^\mathrm{up,bat}$ stands for the battery capacity reserved for the regulation-up service given by $b_t^\mathrm{up,bat} = b_t^\mathrm{up}-b_t^\mathrm{up,pv}$, and $b_t^\mathrm{dn,bat}$ stands for the portion of the regulation-down commitment that must be provided by battery charging and is given by 
\begin{align}
    b_t^\mathrm{dn,bat} =  \max\left\{ b_t^\mathrm{dn} - \left(\kappa P_t^\mathrm{pred} -b_t^\mathrm{up,pv}\right),\, 0 \right\}, 
    \label{eq:regulation_down_battery}
\end{align}
which corresponds to the residual downward AS requirement that cannot be met by the available PV energy.
Within the limits \eqref{eq:energy_POI_limit} and \eqref{eq:energy_feasible}, the energy market bid $b_t^{\mathrm{e}}$ is determined from the agent action $a_t^{\mathrm{e}} \in [0, 1]$, which specifies the desired position within the feasible energy bidding range as follows:
\begin{align}
  b_t^\mathrm{e} = \min\Bigl\{ & \max\bigl\{\tilde{b}_t^\mathrm{e}, \, P_t^\mathrm{pred} - b_t^\mathrm{up,pv} - B_t^{\mathrm{chg}} \bigr\}, \notag \\
  &  P_t^\mathrm{pred} - b_t^\mathrm{up,pv} + B_t^{\mathrm{dis}} \Bigr\}
\end{align}
with $\tilde{b}_t^\mathrm{e}$ given by
\begin{align}
   \tilde{b}_t^\mathrm{e} = & 
   a_t^\mathrm{e}(P_\mathrm{max}^\mathrm{poi} - b_t^\mathrm{res} -b_t^\mathrm{up}) 
   + (1-a_t^\mathrm{e})(P_\mathrm{min}^\mathrm{poi} +b_t^\mathrm{dn}).
   \label{eq:energy_bid}
\end{align}
This decision captures the agent's strategy for energy market participation in anticipation of future price variations. 

\subsubsection{Real-time Battery Control and Imbalances}
\label{sec:imbalances}

In real-time operation, the actual power injection at the POI may differ from $b_t^\mathrm{e}$ due to PV generation uncertainty and AS activations.
In this paper, since the capacity and energy required for the AS provision are explicitly ensured in \secref{sec:AS_provision}, we focus on deviations of the baseline setpoint at POI before AS activations. 
This deviation is attributable to the imbalances in the energy market as described in \secref{sec:energy_market} and does not imply any non-delivery of AS commitments.
In the following, we separately formulate the cases where the scheduled energy market bid exceeds the available power, leading to an energy \emph{shortfall}, and where it is lower than the available power, resulting in energy \emph{surplus}.

We first consider the shortfall case, where the scheduled energy exceeds the available energy for the energy market after accounting for the upward AS provision. 
Specifically, with the actual PV power $p_t^\mathrm{avail}$ at time $t$, an energy shortfall may arise when
\begin{align}
  b_t^\mathrm{e} > p_t^\mathrm{avail} - b_t^\mathrm{up,pv},
\end{align}
where $b_t^\mathrm{up,pv}$ is the reserved PV power for the regulation-up service given in \eref{eq:pv_up_regulation}. 
When $b_t^\mathrm{up,pv}=0$, \ie, 
$b_t^\mathrm{up} \le M_t^{\mathrm{bat}\uparrow}$, 
the shortfall in the energy market can be mitigated by discharging the battery.
The real-time battery discharge for imbalance mitigation can be given by 
\begin{align}
  x_t^\mathrm{disE(short)}
  =
  \min\Bigl\{
  a_t^{\mathrm{imb}} \bigl(b_t^{\mathrm{e}} - p_t^\mathrm{avail}+b_t^\mathrm{up,pv}\bigr),
  \, B_t^\mathrm{dis}
  \Bigr\} \Delta t,  \label{eq:discharge_energy}
\end{align}
where the agent action $a_t^{\mathrm{imb}} \in [0,1]$ is introduced following the idea in \cite{Jeong2023} to allow the agent to determine how battery energy is allocated between mitigating current imbalances and preserving battery energy margin for future arbitrage opportunities.
The energy shortfall in the energy market schedule is given by 
\begin{align}
  \Delta E_t^\mathrm{short} = \left(b_t^\mathrm{e} - p_t^\mathrm{avail} + b_t^\mathrm{up,pv}\right)\Delta t - x_t^\mathrm{disE(short)}. 
  \label{eq:imbalance_short}
\end{align}
Note that when $b_t^\mathrm{up,pv} > 0$, \ie, $b_t^\mathrm{up} > M_t^\mathrm{bat\uparrow}$, it can be shown that $B_t^\mathrm{dis} = 0$, and thus $x_t^\mathrm{disE(short)}=0$.


We next consider the surplus case, where the scheduled energy is lower than the energy available for the energy market, \ie, 
\begin{align}
  b_t^\mathrm{e} \le p_t^\mathrm{avail} - b_t^\mathrm{up,pv}.
\end{align}
A part of this surplus can be absorbed by the battery charging:
\begin{align}
  & x_t^\mathrm{chgE(sur)}
  = 
  \min\left\{
  a_t^\mathrm{imb}
  \bigl(p_t^\mathrm{avail} - b_t^\mathrm{up,pv} - b_t^\mathrm{e}\bigr),\,
  \hat{B}_t^\mathrm{chg}
  \right\}  \Delta t,   \\
  & \hat{B}_t^\mathrm{chg} =  \min\bigl\{ 
      P_\mathrm{bat} - b_t^\mathrm{dn,bat},\, 
     \bigl( E_t^\mathrm{bat\downarrow} -\hat{b}_t^\mathrm{dn,bat}H_\mathrm{dn} \bigr)/\Delta t \bigr\},  
\end{align}
where $\hat{b}_t^\mathrm{dn,bat}$ is defined analogously to $b_t^\mathrm{dn,bat}$ in \eref{eq:regulation_down_battery}, except that the predicted PV output is replaced by the realized value $p_t^\mathrm{avail}$: 
\begin{align}
    \hat{b}_t^\mathrm{dn,bat} = & \max\left\{  b_t^\mathrm{dn} - \left(p_t^\mathrm{avail} -b_t^\mathrm{up,pv}\right),\, 0 \right\}. 
\end{align}
Furthermore, since the power injection at the POI must satisfy the upper bound $P_{\max}^\mathrm{poi}$, it may require reduction of the PV output given by 
\begin{align}
  x_t^\mathrm{cur}
  =
  \max\Bigl\{ 
  p_t^\mathrm{avail}\Delta t - x_t^\mathrm{chgE(sur)} - P_{\max}^\mathrm{poi}\Delta t,\,
  0
  \Bigr\}.
\end{align}
Consequently, the energy surplus in the energy market is given by  
\begin{align}
    \Delta E_t^\mathrm{sur} = (p_t^\mathrm{avail} -b_t^\mathrm{up,pv} -b_t^\mathrm{e})\Delta t  - x_t^\mathrm{chgE(sur)} -x_t^\mathrm{cur}. 
    \label{eq:imbalance_surplus}
\end{align}


Finally, the battery SOC dynamics is described below. 
The SOC evolves according to the actual battery charging and discharging realized in real-time operation, including those induced by AS activations.
Accordingly, the battery SOC update introduced in \eref{eq:soc_dynamics} can be rewritten as
\begin{equation}
  x^{\mathrm{soc}}_{t+1}
    = x^{\mathrm{soc}}_{t}
      + \eta_\mathrm{c} \frac{x_t^\mathrm{chgE} + x_t^\mathrm{chgAS}}{E_\mathrm{bat}}
      - \frac{x_t^\mathrm{disE}+ x_t^\mathrm{disAS}}{\eta_\mathrm{d}E_\mathrm{bat}},
\end{equation}
where $E_t^{\mathrm{chgE}}$ and $E_t^{\mathrm{disE}}$ represent the battery energy charged from and discharged to the energy market, respectively, and given by 
\begin{align}
 x_t^\mathrm{chgE} = 
 \begin{cases}
    0, & b_t^\mathrm{e} > p_t^\mathrm{avail} - b_t^\mathrm{up,pv} \\
    x_t^\mathrm{chgE(sur)}, & b_t^\mathrm{e} \le p_t^\mathrm{avail} - b_t^\mathrm{up,pv} 
 \end{cases}, \label{eq:chgE} \\
 x_t^\mathrm{disE} = 
 \begin{cases}
    x_t^\mathrm{disE(short)},& b_t^\mathrm{e} > p_t^\mathrm{avail} - b_t^\mathrm{up,pv} \\
    0 & b_t^\mathrm{e} \le p_t^\mathrm{avail} - b_t^\mathrm{up,pv}  \label{eq:disE}
 \end{cases}.
\end{align}
The terms $x_t^\mathrm{chgAS}$ and $x_t^{\mathrm{disAS}}$ represent the realized
battery energy charged and discharged due to AS activations,
respectively.
Specifically, the energy charged due to downward AS activation is given by
\begin{align}
  x_t^\mathrm{chgAS} = \min \left\{ b_t^\mathrm{dn}h_t^\mathrm{dn}, \, E_t^\mathrm{bat\downarrow}-x_t^\mathrm{chgE} \right\}, 
  \label{eq:chgAS}
\end{align}
where $h_t^{\mathrm{dn}}$ is the realized duration of downward AS activation during period~$t$.
The battery energy discharged due to upward AS activations is given by
\begin{align}
  x_t^\mathrm{disAS} = p_t^\mathrm{res}h_t^\mathrm{res} + (b_t^\mathrm{up} - b_t^\mathrm{up,pv}) h_t^\mathrm{up}. 
\end{align}
where $h_t^\mathrm{res}$ and $h_t^\mathrm{up}$ represent the realized activation durations of contingency reserve and regulation-up services, respectively.


\subsection{Design Optimization Component} 
\label{sec:design_update}

In the proposed co-optimization framework, the design parameter $\omega$ is regarded as an additional action that remains fixed throughout each episode and is generated by a state-independent stochastic policy with distribution $p_\mu(\omega)$.
Here, the learnable parameter $\mu$  represents the current belief about which system designs are likely to be effective.
In this study, the design vector is defined in \eref{eq:design_variable} and  the corresponding mean parameter is given by
\begin{align}
    \mu = \left\{ \mu_{P_\mathrm{pv}},\; \mu_{E_\mathrm{bat}},\; \mu_{P_\mathrm{bat}} \right\}.
\end{align}
Following the idea introduced in~\cite{Mantani2025:TEMPR}, we model the design distribution as a multivariate Gaussian
\begin{align}
    p_\mu(\omega) = \mathcal{N}(\mu,\sigma^2 I),
\end{align}
where $\sigma^2$ is a fixed variance controlling exploration in the design space.

Let $G$ denote the episode return obtained under a sampled design $\omega$.
In this paper, the return represents the economic value of the hybrid resource over the episode horizon and is defined as
\begin{align}
    G
    =
    W_\mathrm{anu}\sum_t (F_t^\mathrm{e}+F_t^\mathrm{as}) + F_\mathrm{cap}
    -
    C_\mathrm{cap}(\omega) (1-\eta),
    \label{eq:net_profit_G}
\end{align}
where the parameter $\eta \in [0,1]$ is introduced to improve the convergence of the co-optimization process~\cite{Mantani2025:TEMPR}.
During training, $\eta$ is scheduled to gradually decrease from 1 to 0 so that the influence of the capital cost is progressively introduced into the optimization.
Following the policy gradient theorem~\cite{Sutton2018:RL}, the gradient of the expected return with respect to $\mu$ is given by  
\begin{align} \label{eq:gradient_mu}
 \nabla_\mu \mE[G] = \mE \left[ \nabla_\mu \ln p_\mu(\omega) G \right]. 
\end{align}
For the Gaussian design distribution $p_\mu(\omega)=\mathcal{N}(\mu,\sigma^2 I)$, the gradient of the log-likelihood is given by
\begin{align}
\nabla_\mu \ln p_\mu(\omega)
=
\frac{\omega-\mu}{\sigma^2}.
\end{align}
Substituting this into~\eqref{eq:gradient_mu} yields
\begin{align}
\nabla_\mu \mathbb{E}[G]
=
\mathbb{E}
\left[
\frac{\omega-\mu}{\sigma^2} G
\right].
\end{align}
In practice, this expectation is approximated using a batch of recent episodes.
Let $\{\omega_i,G_i\}$ denote the sampled design and the corresponding return for episode $i$.
Using a sample mean baseline $\overline{G}_{\mathcal E}$ to reduce variance, the empirical estimator of the gradient becomes
\begin{align}
\nabla_\mu \mathbb{E}[G]
\approx
\frac{1}{N_\mathrm{up}}
\sum_{i\in\mathcal E}
\frac{\omega_i-\mu}{\sigma^2}
\left(G_i- \overline{G}_{\mathcal E}\right),
\label{eq:gradient_G}
\end{align}
where $\mathcal E$ denotes the indices of the most recent $N_\mathrm{up}$ episodes.
Since $\omega$ is a vector design parameter, the gradient in~\eqref{eq:gradient_G} is interpreted component-wise, yielding the update direction for each design parameter.
The mean parameter $\mu$ is then updated using a gradient ascent step
\begin{align}
\mu
\leftarrow
\mu
+
\alpha_\mu
\nabla_\mu \mathbb{E}[G],
\end{align}
where $\alpha_\mu$ represents the learning rate for the design parameters.
This update can be interpreted as a stochastic policy gradient method in the design space.
By sampling designs and updating the distribution mean according to the observed returns, the method gradually shifts the design distribution toward more profitable system configurations.

\section{Numerical Results}
\label{sec:simulation}

This section demonstrates the effectiveness of the proposed co-design framework for PV-battery hybrid resources. 

\subsection{Simulation Settings}
\label{sec:settings}

The simulations are conducted using time-series data of electricity market prices and PV generation based on historical CAISO data.
To obtain representative system-level signals while avoiding location-specific bias, the price data are computed as the average across all nodes in the CAISO network.
Similarly, the solar generation profile is obtained by averaging across all zones and then  scaled to match the capacity level of the target system.
All time-series data are resampled to a uniform resolution of $\Delta t = \SI{1}{h}$.
The resulting dataset provides a synthetic yet realistic market conditions, enabling the analysis to focus on structural characteristics of hybrid resource design and operation. 
For the analyses in \secref{sec:comparison} and \secref{sec:reconfiguration}, a one-week period (the first week of July 2022) is used for detailed examination of operational behaviors and design trade-offs.
In \secref{sec:long_term}, the proposed framework is further evaluated using one-year data to assess long-term performance and scalability.

\begin{table}[t!]
\centering
\renewcommand{\arraystretch}{1.05}
\caption{Parameters for numerical simulation}
\label{tab:parameter}
\begin{tabular}{|c|c|c|}
\hline
 Symbol & Parameter & Values \\
 \hline
$S_{\min}$ & Minimum value of SOC  & $0.1$ \\
$S_{\max}$ & Maximum value of SOC  & $0.9$ \\
$\eta_\mathrm{c}$ &  Charging efficiency  & $0.95$ \\
$\eta_\mathrm{d}$ &  Discharging efficiency  & $0.95$ \\
$\beta_\mathrm{deg}$ &  Battery degradation cost  & $\mathrm{\$1.0/MWh}$ \\
$\pi_t^\mathrm{imb}$ &  Penalty factor for imbalances  &  $1.0$ \\
$\phi_\mathrm{pv}$ & Firm PV capacity factor & $0.4$ \\
$H_\mathrm{cr}$ & Capacity duration requirement & $\SI{4.0}{h}$\\
$\kappa$ & Reliability factor for AS provision & $0.7$ \\
$H_\mathrm{res}$ & Contingency reserve requirement & $\SI{0.5}{h}$ \\
$H_\mathrm{up}$ & Up-regulation requirement & $\SI{0.35}{h}$ \\
$H_\mathrm{dn}$ & Down-regulation requirement & $\SI{0.35}{h}$ \\
 \hline
 \end{tabular}
\end{table}

We consider a PV-battery hybrid resource connected to the grid, where the AC-side capacity at the POI is fixed to $P^\mathrm{poi}_{\max} = 10~\mathrm{MW}$, and the design problem focuses on the optimal sizing and operation of the DC-side components.
This setting reflects practical interconnection limits, where expanding the POI capacity is often costly or infeasible, and thus optimal utilization of limited interconnection capacity is critical.
The PV system cost is assumed to be $1080$\,\$/kW based on~\cite{Seel2024} and is annualized over 20 years.
For the battery system, the cost parameters are assumed to be 241\,\$/kWh for energy capacity and 372\,\$/kW for power capacity based on~\cite{cole2021cost}. 
The battery cost is annualized assuming a 5-year lifetime, except in the one-week case studies, where a 7-year lifetime is assumed so that the investment cost remains commensurate with the revenue obtained over the evaluation horizon. 
The capacity payment is modeled based on the Resource Adequacy price reported in~\cite{CPUC_RA_2022}, which is set to 8.31\,\$/kW-month.
Other system parameters are summarized in \tabref{tab:parameter}.

The proposed co-design framework is implemented in Python using PyTorch. 
For the DDPG agent, both the actor and critic networks consist of three hidden layers with 64 units each, and the learning rates for both networks are set to $1\times10^{-4}$  in the experiments of \secref{sec:comparison} and \secref{sec:reconfiguration}.
The LSTM network used for price and solar generation forecasting has a hidden layer size of 64 units.
The learning rate for the design parameters is set to $\alpha_\mu = 1\times 10^{-5}$.
The learning rate for the design parameters is also adjusted to $\alpha_\mu = 1\times 10^{-6}$.
In \secref{sec:long_term}, which considers long-term training over one-year data, the network size is increased to 128 units for both actor and critic networks, and the learning rates are reduced to $1\times10^{-7}$ for stable training.
The learning rate for the design parameters is also adjusted to $\alpha_\mu = 5\times 10^{-7}$. 
To ensure stable training, normalization of state variables and reward scaling are applied.
All simulations are performed on a workstation equipped with an AMD EPYC 7763 CPU and an NVIDIA RTX 6000 Ada GPU, and the reported results are obtained after convergence of the training process.
Our implementation is available at~\url{https://github.com/uhyogo-epa/HROpt}.

\subsection{Hybrid vs Co-located Resources}
\label{sec:comparison}

In this subsection, we compare the operational economic performance of hybrid and co-located resources under fixed system sizes.
The PV capacity is set to $P_\mathrm{pv}=\SI{11}{MW}$, corresponding to a 10\% oversizing. 
The battery system is fixed at $P_\mathrm{bat}=\SI{5}{MW}$ and $E_\mathrm{bat}=\SI{20}{MWh}$, which satisfies the 4-hour rule in capacity remuneration. 
For the co-located configuration, the corresponding formulation is provided in Appendix~\ref{sec:co-located}.
\Figref{fig:learning_curve} shows the learning progress in terms of episode reward.
 A total of 10 runs with different random seeds were conducted for each configuration. For the hybrid case, one run resulted in significantly lower rewards and was excluded.
The results indicate that the hybrid resource consistently achieves higher rewards than the co-located resource, demonstrating its superior operational economic performance under the same system size.

\begin{figure}
    \centering
    \includegraphics[width=0.90\linewidth]{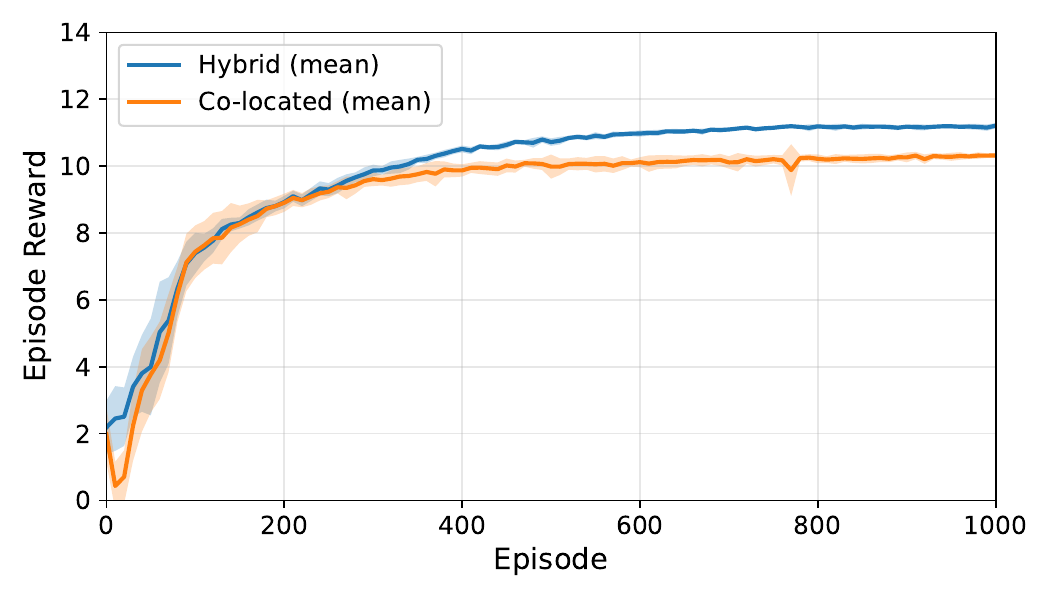}
    \caption{Progress of episode rewards during training}
    \label{fig:learning_curve}
\end{figure}

\begin{figure}
    \centering
    \includegraphics[width=0.90\linewidth]{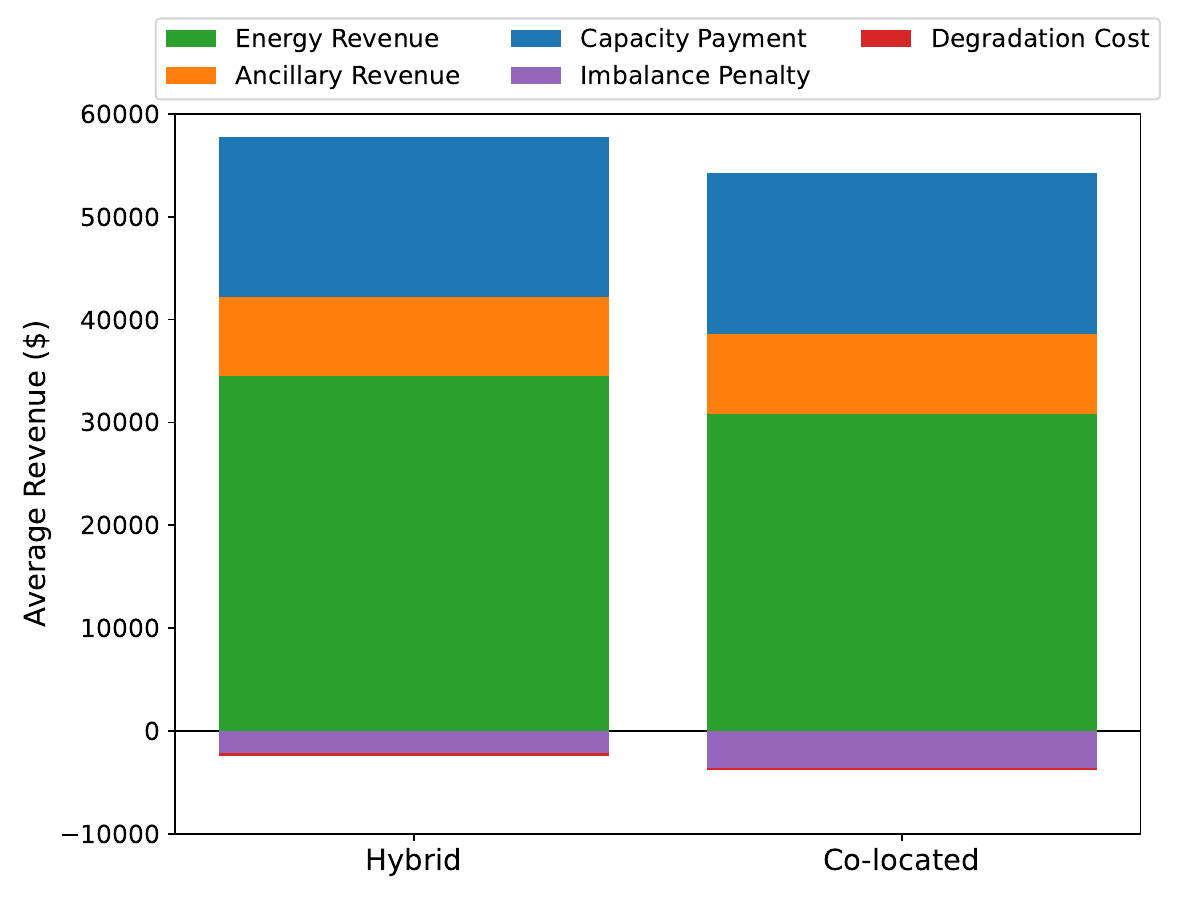}
    \caption{Breakdown of revenue components of hybrid and co-located resources}
    \label{fig:revenue_breakdown}
\end{figure}

\begin{figure}
\centering
\begin{minipage}{1.0\columnwidth}
    \centering
     \includegraphics[width=0.98\columnwidth]{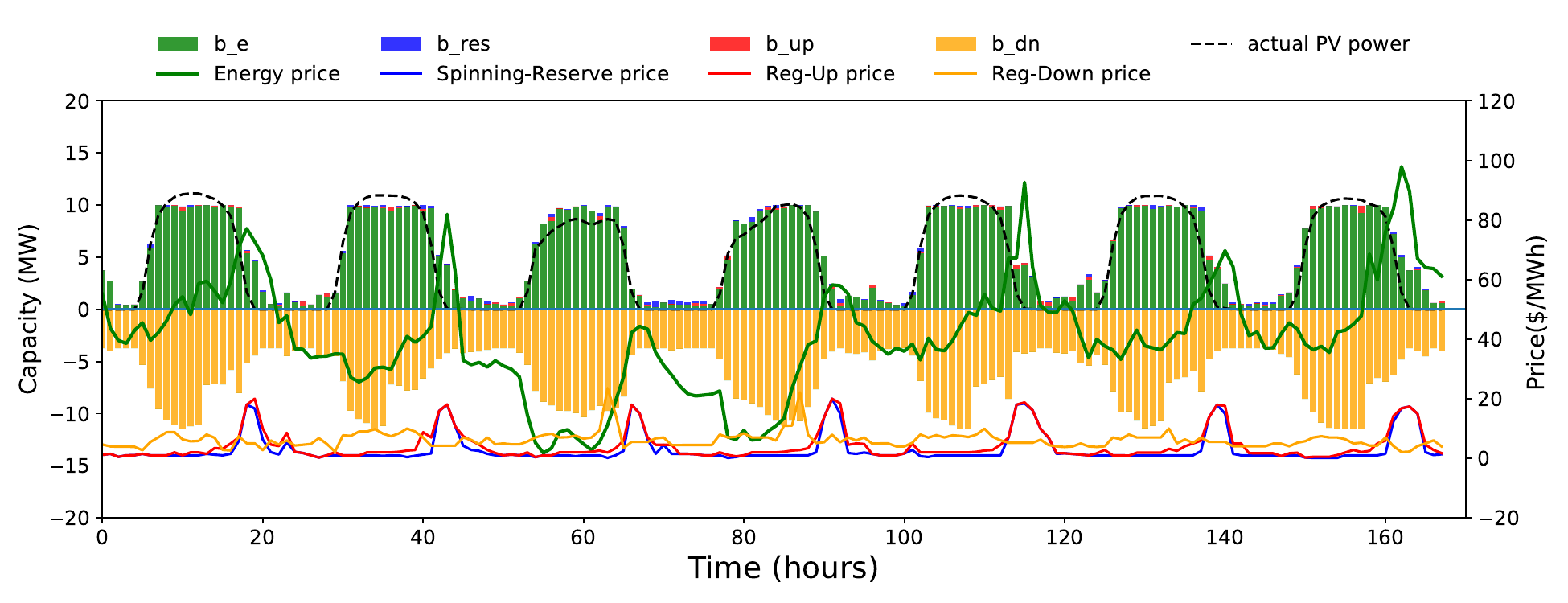}
     \subcaption{Hybrid resource} \label{fig:comparison_hybrid}
    \label{fig:solar_generation}
\end{minipage}
\begin{minipage}{1.0\columnwidth}
    \centering
    \includegraphics[width=0.98\columnwidth]{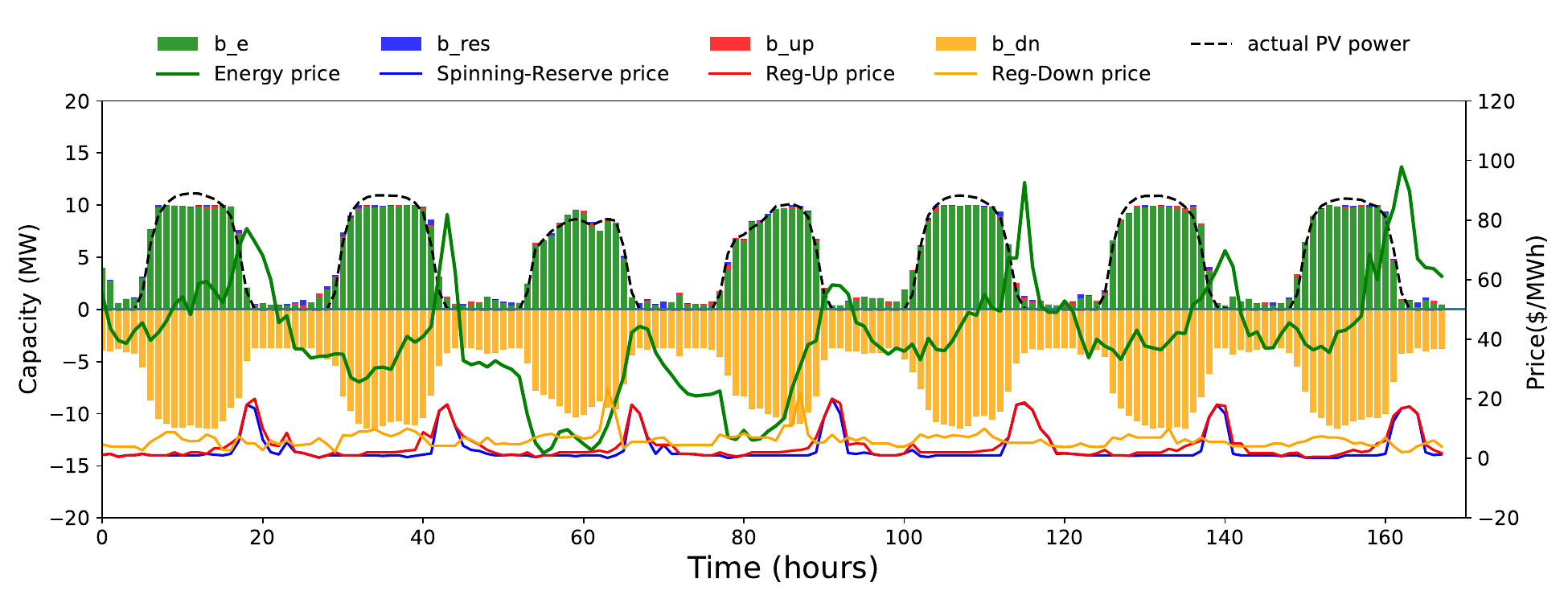}
     \subcaption{Co-located resource}
    \label{fig:comparison_colocated}
\end{minipage}
\caption{Time-series behaviors of hybrid and co-located resources}
\label{fig:time_series_comparison}
\end{figure}

\Figref{fig:revenue_breakdown} presents the breakdown of the revenue components, including energy market revenue (the first term in \eref{eq:energy_penalty_model}), AS revenue, capacity payment, battery degradation cost, and imbalance penalties (the second term in \eref{eq:energy_penalty_model}).
The results show that the imbalance penalties are smaller in the hybrid resource, amounting to \$2193 compared to \$3622 for the co-located resource.
This is because the hybrid resource can internally compensate for real-time PV forecast errors using the battery, which corresponds to the second advantage discussed in \secref{sec:hybrid_benefits}.
Regarding the allocation of market participation, the hybrid resource obtains a larger portion of its revenue from the energy market, whereas the co-located resource relies more on AS markets.
To gain further insight into these operational differences, \figref{fig:time_series_comparison} illustrates the time-series behaviors of the hybrid and co-located resources.
As discussed in \secref{sec:hybrid_benefits}, one of the key advantages of hybrid resources is the ability to recover clipped PV energy through DC coupling.
In the hybrid configuration, PV generation exceeding the POI limit can be directly stored in the battery and later dispatched during high-price periods.
This behavior is observed in \figref{fig:comparison_hybrid}, where energy delivery is increased during evening hours compared with \figref{fig:comparison_colocated}. 
Instead, in the co-located resource, the available charging capacity tends to be allocated to regulation-down services, leading to higher AS revenue.
We also examined the impact of PV oversizing ratios.
As the oversizing ratio increases, the amount of clipped PV energy that can be recovered and shifted via the battery becomes larger, leading to higher energy market revenue.
At the same time, the benefit of imbalance penalty reduction becomes less pronounced, because larger oversizing results in more frequent operation at the POI limit, effectively capping the scheduled energy at the maximum deliverable output and thereby reducing the sensitivity to PV forecast errors.

Overall, the proposed framework enables a quantitative comparison of hybrid and co-located resources by explicitly capturing multi-market trade-offs and real-time imbalance mitigation.
Such analysis has not been possible in existing studies that focus only on energy markets~\cite{Gomez2017,Bhattacharjee2025} or assume perfect renewable forecasts~\cite{Huang2021}.


\subsection{Co-optimization of Hybrid Resources}
\label{sec:reconfiguration}

\begin{figure}
    \centering
    \includegraphics[width=0.90\linewidth]{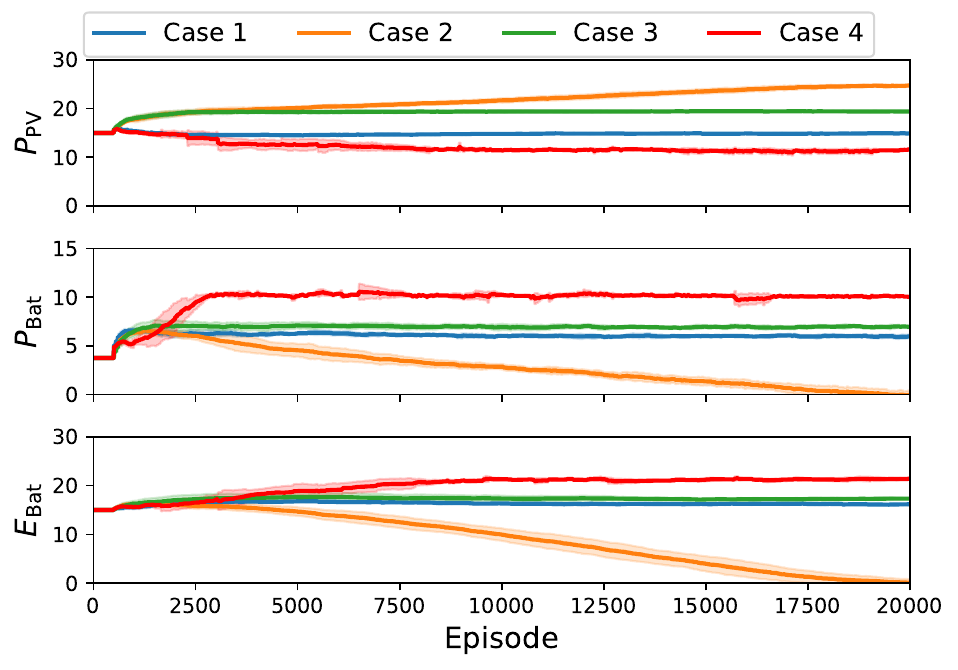}
    \caption{Evolution of the design parameters during training.}
    \label{fig:mu_progress}
\end{figure}

\begin{figure}
    \centering
    \includegraphics[width=0.95\linewidth]{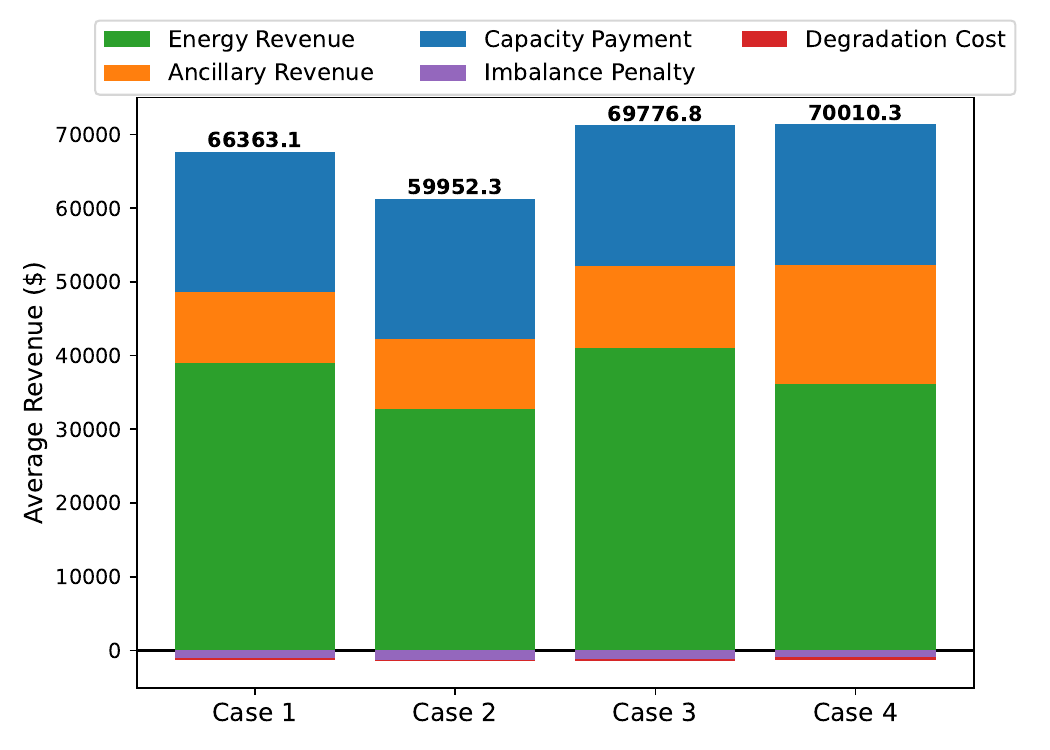}
    \caption{Breakdown of revenue components of hybrid and co-located resources}
    \label{fig:revenue_co_optimization}
\end{figure}

\begin{figure}
\centering
\begin{minipage}[b]{1.0\columnwidth}
    \centering
     \includegraphics[width=0.95\columnwidth]{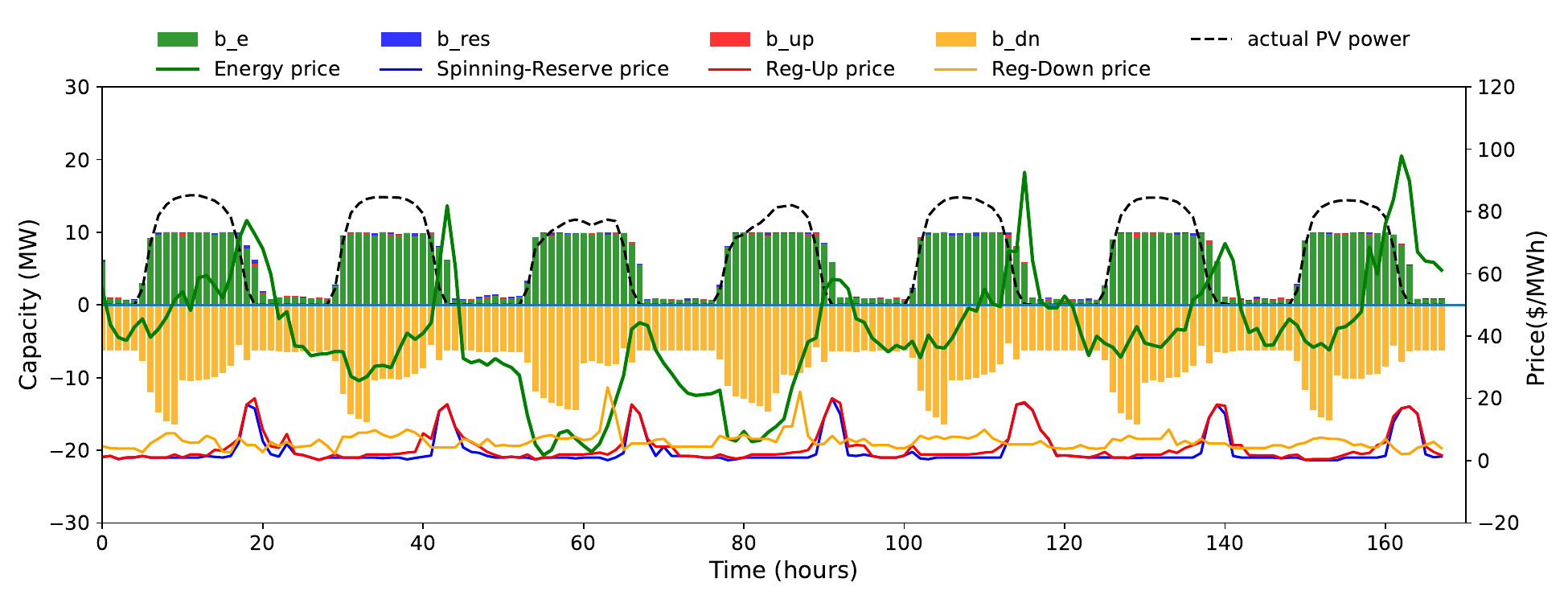}
     \subcaption{Case 1: Baseline}
\end{minipage}
\begin{minipage}[b]{1.0\columnwidth}
    \centering
    \includegraphics[width=0.95\columnwidth]{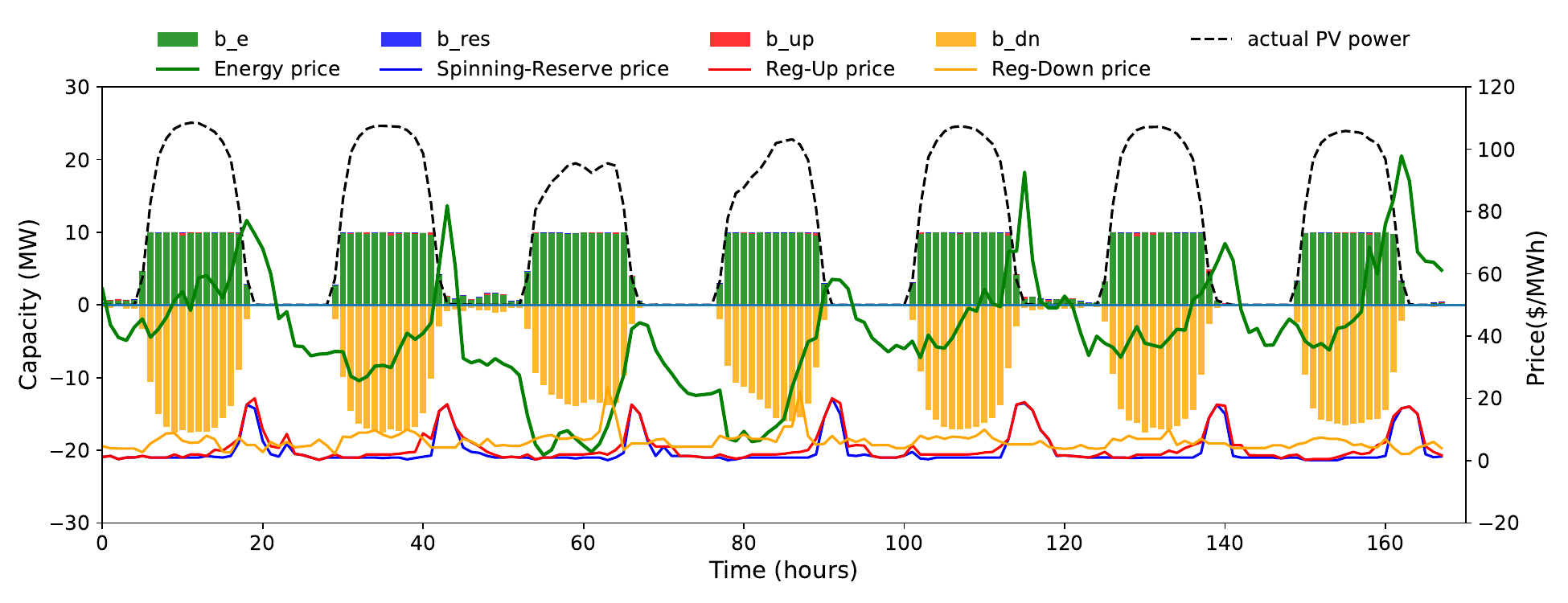}
     \subcaption{Case 2: Capacity market reform}
\end{minipage}
\begin{minipage}[b]{1.0\columnwidth}
    \centering
    \includegraphics[width=0.95\columnwidth]{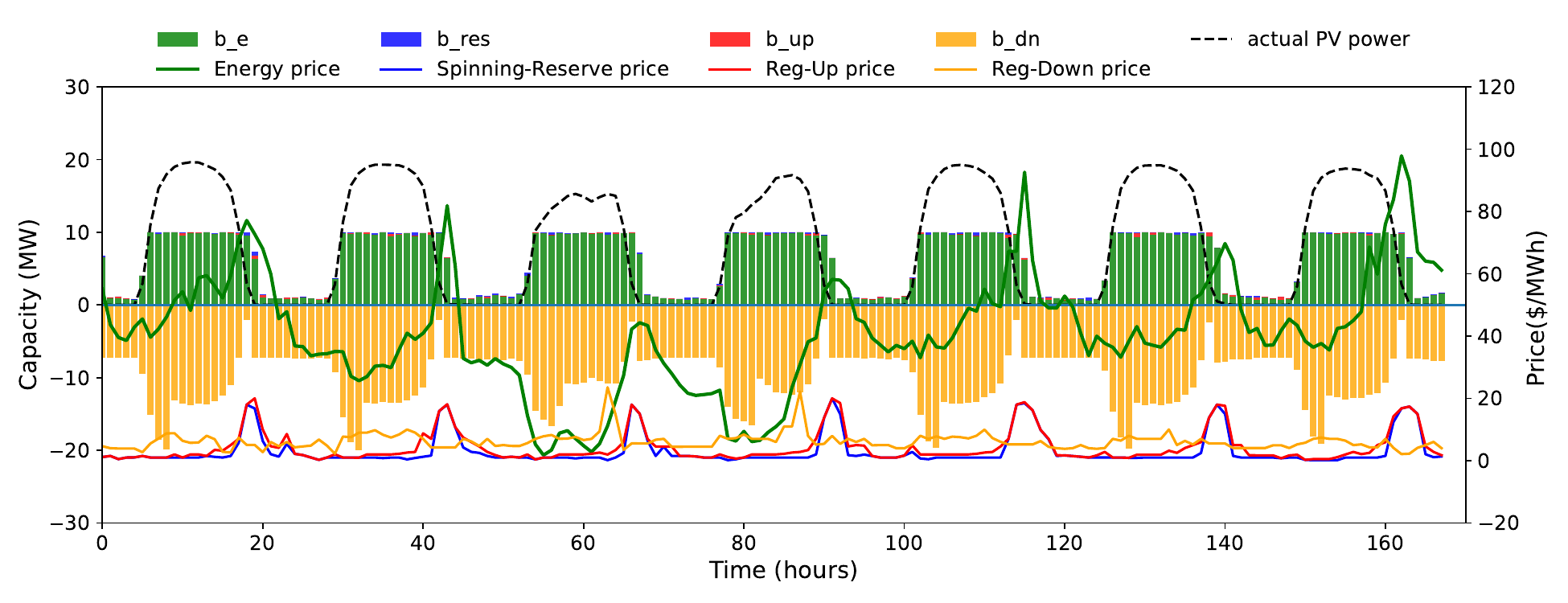}
     \subcaption{Case 3: Battery cost reduction}
\end{minipage}
\begin{minipage}[b]{1.0\columnwidth}
    \centering
    \includegraphics[width=0.95\columnwidth]{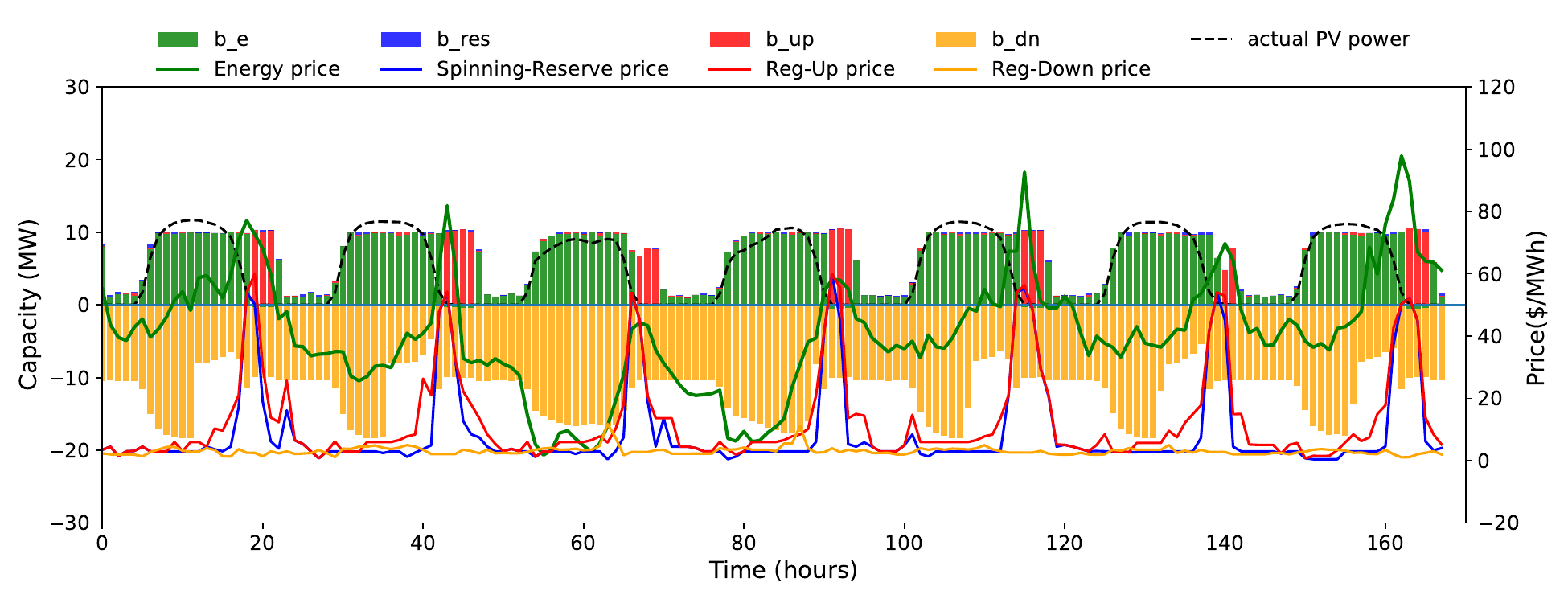}
     \subcaption{Case 4: AS price increase}
\end{minipage}
\caption{Time-series behaviors under hypothetical scenarios}
\label{fig:time_series_co_optimization}
\end{figure}

This subsection demonstrates the effectiveness of the proposed co-optimization framework by jointly optimizing system design and operational strategies under different market and policy scenarios.
We consider four hypothetical scenarios to investigate how optimal system configurations and bidding strategies adapt to changing environments: 
\begin{enumerate}[label=(Case\arabic*), leftmargin=4.0em, itemsep=0pt, topsep=2pt]
    \item \textit{Baseline:}  Same system and market settings as in \secref{sec:comparison}, with optimized system design.
    \item \textit{Capacity market reform:} Increase in the minimum duration requirement for capacity remuneration from $H_{\mathrm{cr}} = 4$ h to 8 h.
    \item \textit{Battery cost reduction:} Scaling of battery cost parameters by a factor of 0.9 from Case\,2, with all other settings unchanged.
    \item \textit{AS price variation:} Scaling of contingency reserve and regulation-up prices by a factor of 3.0 and regulation-down price by a factor of 0.5 from Case\,1, with all other settings unchanged.
\end{enumerate}
The results are summarized in \tabref{tab:co_optimization}, together with the evolution of the design parameters during training shown in \figref{fig:mu_progress}, the breakdown of revenues in \figref{fig:revenue_co_optimization}, and representative time-series operation in \figref{fig:time_series_co_optimization}. 
Due to the nonconvex nature of the co-optimization problem, multiple locally optimal solutions may emerge. In this study, ten independent runs with different random seeds are performed for each case, and a representative solution is selected based on both its frequency of occurrence and its economic performance. Specifically, the reported solution corresponds to the most frequently observed pattern, which also achieves the highest net profit among the obtained solutions. This solution is reproduced in 6, 10, 9, and 5 out of 10 runs for Cases 1–4, respectively.

\begin{table}[t]
\centering
\caption{Optimal sizing results under different scenarios}
\label{tab:co_optimization}
\begin{tabular}{lcccc}
\hline
 & Case 1 & Case 2 & Case 3 & Case 4 \\
\hline
\multicolumn{5}{l}{\textit{Design}} \\
$P_{\mathrm{pv}}$ (MW) & 14.92 & 24.77 & 19.41 & 11.53 \\
$P_{\mathrm{bat}}$ (MW) & 5.98 & 0 & 6.94 & 9.99 \\
$E_{\mathrm{bat}}$ (MWh) & 16.22 & 0 & 17.33 & 21.42 \\
Duration (h) & 2.71 & 0 & 2.50 & 2.14 \\
\hline
\multicolumn{5}{l}{\textit{Annual Cost}  (k\$) } \\
CAPEX & 1682 & 1338 & 1917 & 1891 \\
\hline
\multicolumn{5}{l}{\textit{Annual Revenue} (k\$)} \\
Energy market    & 1983 & 1641 & 2078 & 1838 \\
AS market         & 499 &  494 &  586 &  844 \\
Capacity payment  & 995 &  991 &  991 &  993 \\
\hline
\multicolumn{5}{l}{\textit{Profit} (k\$) } \\
Total net profit  & 1794 & 1788 & 1738 & 1784 \\
\hline
\end{tabular}
\end{table}

Compared with the fixed-design setting in \secref{sec:comparison}, Case~1 exhibits larger design parameters, particularly in PV capacity and battery energy.
As a result, the total market revenue increases from \$55{,}306.6 in the fixed-design case (corresponding to an annualized value of approximately 2884 k\$) to \$66{,}363.1 (3460 k\$ annualized) in Case 1. 
This improvement demonstrates the importance of jointly optimizing system design and operational strategies, as the co-optimization framework identifies a configuration that more effectively exploits multi-market revenue opportunities.

In this baseline case, the optimal design results in a battery with a duration of  2.71 h, which is below the capacity market requirement. 
In Case 2,  the impact of a policy change that increases the required discharge duration for capacity remuneration is considered, motivated by recent discussions on long-duration storage requirements~\cite{Denholm2023}. 
Under this setting, the effective capacity contribution of the battery is reduced due to the stricter duration requirement.
As a result, in Case 2, the revenue from the capacity market decreases, leading to a reduction in the optimal battery capacity.
Instead, the optimal design shifts toward larger PV capacity.
This result indicates that strengthening duration requirements alone does not necessarily promote battery deployment in hybrid resources.

Case 3 further investigates this policy setting under a reduced battery cost.
Although the duration requirement remains at $H_{\mathrm{cr}} = \SI{8}{h}$, lowering the battery cost makes its deployment economically viable again. 
As a result, the optimal configuration includes a nonzero battery capacity, in contrast to Case 2 where the battery is not installed. However, the resulting battery duration remains around \SI{2.50}{h}. 
This result suggests that, for hybrid resources, the incentive provided by the capacity market reform is not sufficient to promote long-duration storage, even under reduced battery costs, and the economic value of the battery is primarily derived from energy and AS markets. 
At the same time, the result highlights the flexibility of hybrid resource design, which allows the system configuration to adapt to changing market and policy conditions and capture value from alternative revenue streams.

Finally, the impact of AS price signals is examined in Case 4, where the prices of contingency reserves and regulation-up services are increased, while the regulation-down price is reduced. 
Under this condition, AS provision becomes more economically attractive, particularly for services that require upward flexibility. As a result, the optimal design shifts toward a larger battery power capacity, enabling increased AS provision. In contrast, the battery energy capacity increases more moderately, leading to a shorter effective duration compared to Cases 1 and 2. 
The corresponding time-series behavior further illustrates this shift, where higher regulation-up capacities are maintained. 
These results highlight that price signals in AS markets can influence not only operational strategies but also the power-to-energy configuration of the battery system.

\subsection{Applicability to Long-term Data}
\label{sec:long_term} 

This subsection shows the applicability of the proposed framework to long-term operation using one-year time-series data.
To handle the increased computational burden associated with long-term data, a distributed RL framework as in \cite{Horgan2018:Ape-X} is used.
Specifically, the learner and workers are decoupled, and multiple workers are used to collect experience in parallel.
Each worker is assigned one month of data, resulting in 12 parallel workers covering the entire year.
This distributed structure significantly reduces the wall-clock time required for training. 
While the analyses in \secref{sec:comparison} and \ref{sec:reconfiguration}, based on one-week data, required approximately 4 hours, the full-year optimization can be completed in approximately 10 hours using the proposed parallelization scheme. 
Although the full-year dataset spans 52 weeks of operation, the use of parallel workers enables efficient experience collection across different time segments, allowing the optimization to remain computationally tractable. 
This makes it feasible to incorporate long-term temporal variability into the co-design problem.

\begin{figure}
    \centering
    \includegraphics[width=0.95\linewidth]{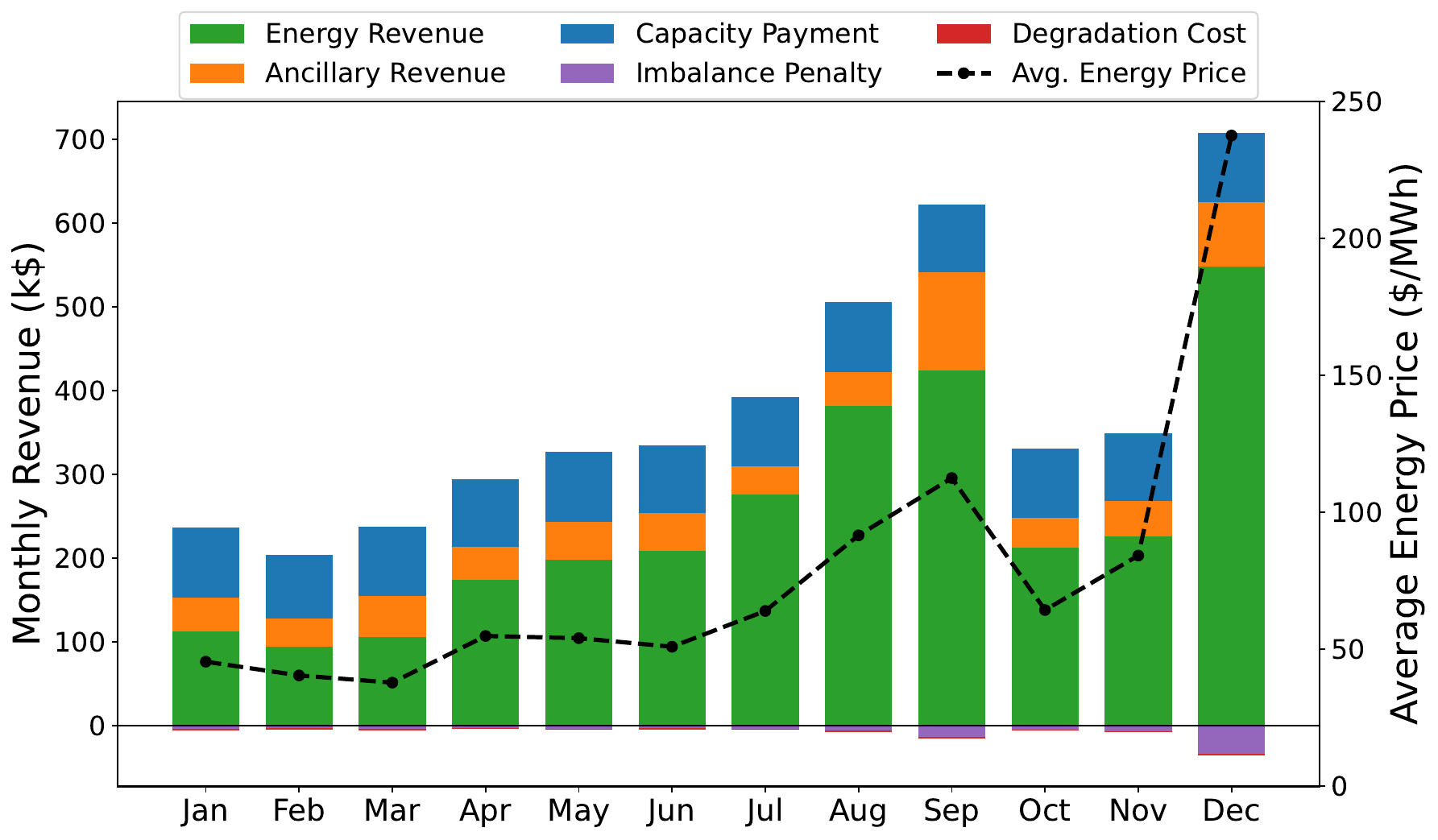}
    \caption{Breakdown of revenue components over the year}
    \label{fig:monthly_breakdown}
\end{figure}

\begin{figure}
\centering
\begin{minipage}[b]{1.0\columnwidth}
    \centering
     \includegraphics[width=0.95\columnwidth]{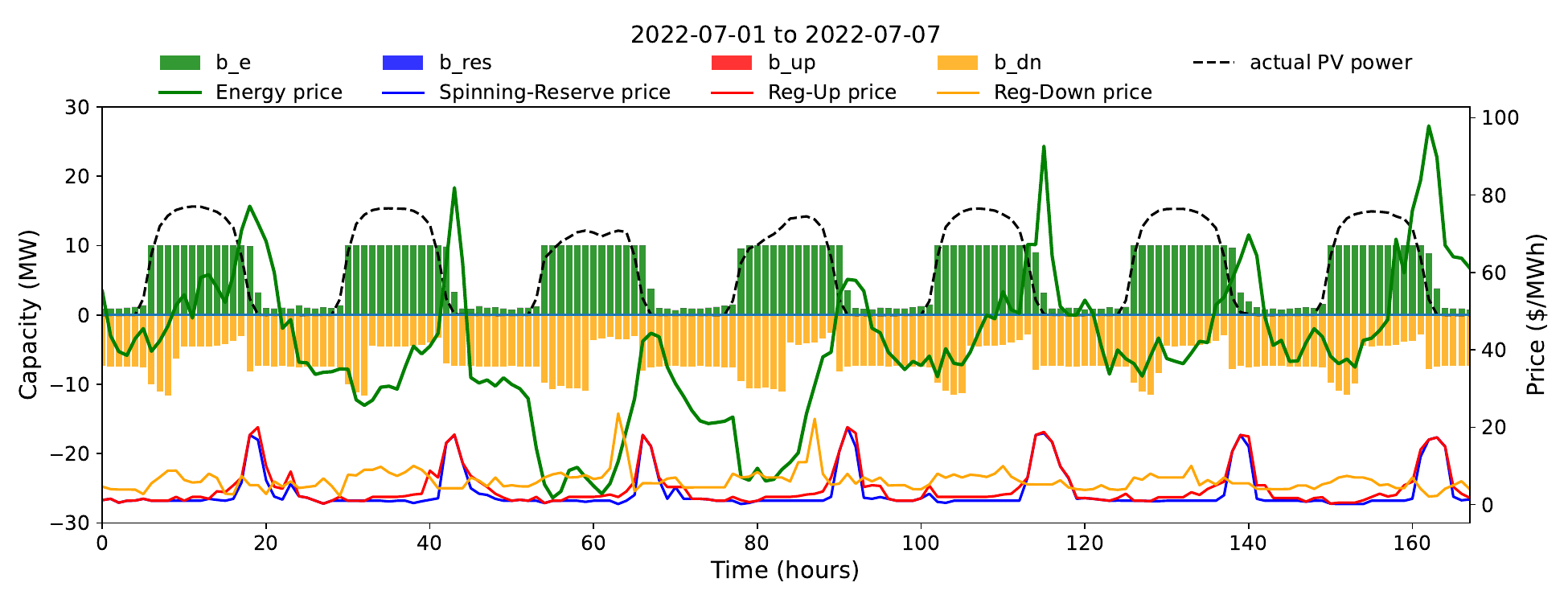}
     \subcaption{1st week of July}
\end{minipage}
\begin{minipage}[b]{1.0\columnwidth}
    \centering
    \includegraphics[width=0.95\columnwidth]{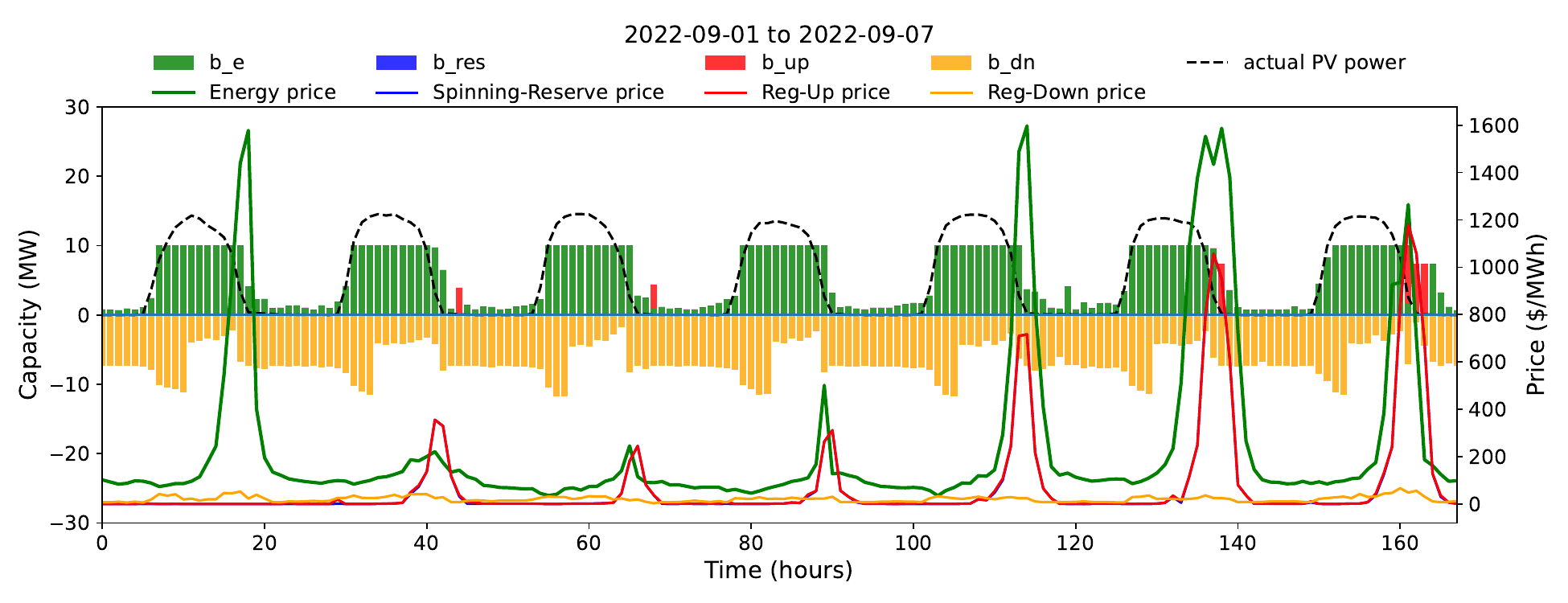}
     \subcaption{1st week of September}
\end{minipage}
\begin{minipage}[b]{1.0\columnwidth}
    \centering
    \includegraphics[width=0.95\columnwidth]{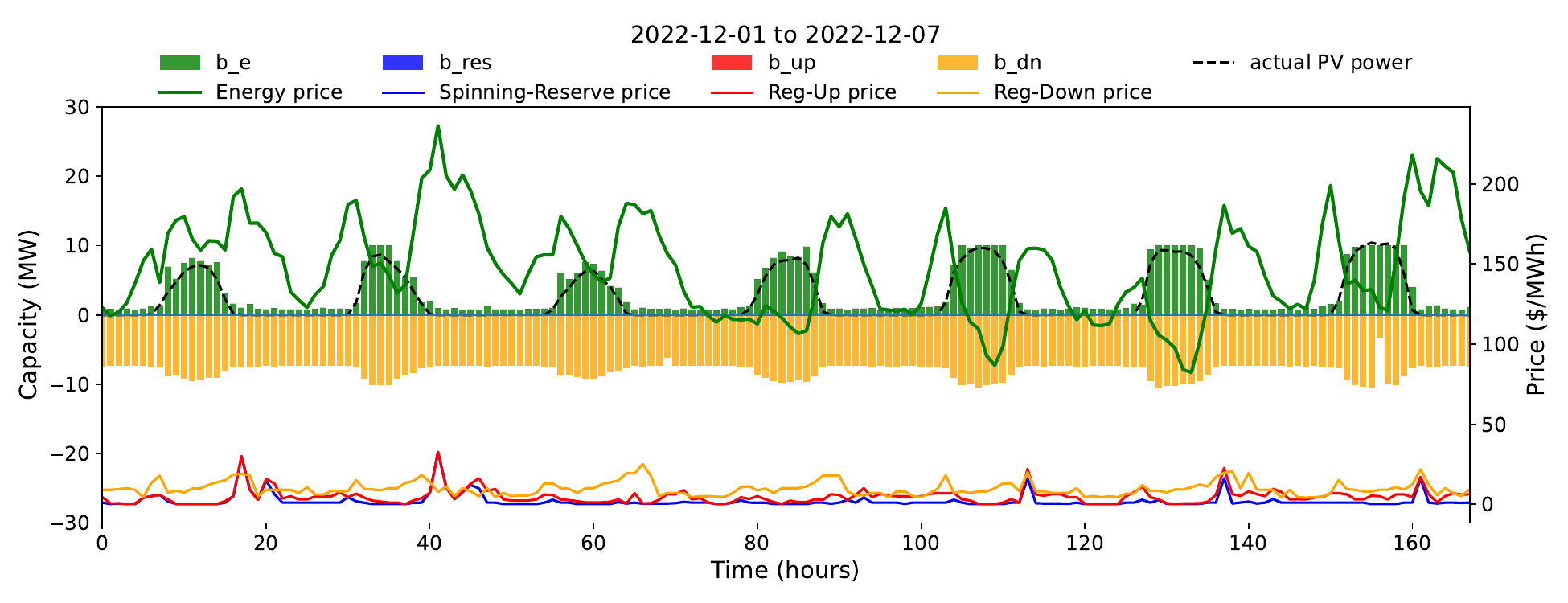}
     \subcaption{1st week of December}
\end{minipage}
\caption{Representative time series in selected months}
\label{fig:time_series_long_data}
\end{figure}

\Figref{fig:monthly_breakdown} presents the results of the full-year analysis. 
The optimal system configuration is determined as a PV capacity of 15.3 MW, a battery power capacity of 9.87 MW, and a battery energy capacity of 14.82 MWh.
As shown in \figref{fig:monthly_breakdown}, seasonal variations in market prices and PV generation across different months lead to different revenue structures, indicating that both operational strategies and economic performance vary significantly over the year.
\Figref{fig:time_series_long_data} presents representative time-series behaviors for selected periods: July as a typical month, September with high ancillary service activity, and December with high market prices.
In September, the results show increased participation in regulation-up services, reflecting favorable price signals for upward flexibility.
In contrast, December exhibits relatively lower energy dispatch volumes, while achieving high revenues due to elevated energy prices.
These findings highlight the importance of considering long-term temporal variability in both system design and operation.
The proposed co-optimization with distributed RL enables such analysis with a practical computational cost, making it suitable for real-world planning problems.

\section{Conclusions}
\label{sec:conclusions}

This paper proposed a DRL-based co-optimization framework for PV-battery hybrid resources, integrating system design and multi-market operational strategies under uncertainty.
The proposed method enables coordinated participation in energy, AS, and capacity markets while explicitly accounting for operational constraints and stochastic variations in prices and renewable generation.
Numerical results demonstrated that the framework can effectively identify economically rational system configurations and operational policies, highlighting the advantages of hybrid resources over co-located alternatives.
Furthermore, the applicability to long-term datasets was validated, showing that seasonal variations significantly influence both revenue structures and optimal operational strategies, while the distributed RL framework enables such analysis with practical computational cost.

Future work includes extending the framework to settings involving multiple decision-makers, such as integrated resource planning and the optimal design and operation of virtual power plants, where coordination among heterogeneous agents and market participants becomes essential.

\appendix
\section{Co-located Resource Formulation}
\label{sec:co-located}

This appendix presents the co-optimization formulation for co-located resources used for the comparative study in \secref{sec:simulation}. 
The design variable $\omega$ follows the same definition as in \eref{eq:design_variable}, with the only difference that $P_{\mathrm{bat}}$ corresponds to the rated power of the battery inverter in the AC-coupled architecture (\figref{fig:ac_coupled_colocated}).
The PV inverter capacity is denoted by $P_\mathrm{inv}$, which limits the PV output as
\begin{equation}
    0 \le p_t^\mathrm{pv} \le P_\mathrm{inv}.
    \label{eq:pv_inverter_limit}
\end{equation}
In this study, we assume $P_\mathrm{inv} = P_\mathrm{max}^\mathrm{poi}$.
The POI power limit \eqref{eq:poi_limit}, the power balance \eqref{eq:dc_power_balance}, and all other battery-related constraints \eqref{eq:converter_limit_chg}--\eqref{eq:soc_limit} remain unchanged. 
Regarding market participation, the market model in \secref{sec:market_model} is applied without modification, except that the PV and the battery are modeled as separate resources with independently submitted bids.
For a fair comparison, the observation, action, and reward definitions in \secref{sec:state_space} are kept identical to those of the hybrid case. 
Ancillary-service bidding capacities are determined using the same procedure as in \secref{sec:AS_provision}, subject to the additional PV inverter constraint \eqref{eq:pv_inverter_limit}.
Specifically, the PV-based AS capacity $M_t^{\mathrm{pv}}$ in \eref{eq:pv_for_AS} is modified as follows: 
\begin{align}
  M_t^\mathrm{pv} &= \min\left\{ 
   \kappa P_t^\mathrm{pred},\, 
   P_\mathrm{inv}
  \right\}. 
\end{align}

For the allocation of AS bids between the PV and the battery, the following rules are applied:
contingency reserves are provided exclusively by the battery;
for regulation-up services, the PV contribution is given by $b_t^\mathrm{up,pv}$ in \eref{eq:pv_up_regulation}, with the remaining capacity allocated to the battery;
for regulation-down services, $b_t^\mathrm{dn,bat}$ in \eref{eq:regulation_down_battery} is allocated to the battery, and the remaining capacity is allocated to the PV.
For the energy market, the PV bid is set to the predicted PV output reduced by the capacity that must be reserved for regulation-up service: 
\begin{align}
 b_t^\mathrm{e,pv}
  =
 p_t^\mathrm{pred} - b_t^\mathrm{up,pv}. 
\end{align}
The PV imbalance arises from the deviation between the predicted and realized PV outputs and is given by
\begin{align}
    \Delta E_t^\mathrm{pv} = \left( \min\bigl\{p_t^\mathrm{avail}, P_\mathrm{inv}\bigr\} - \min\bigl\{p_t^\mathrm{pred}, P_\mathrm{inv}\bigr\}\right)\Delta t. 
\end{align}
The battery energy-market bid $b_t^\mathrm{e,bat}$ is obtained by clipping the residual bid
$\tilde b_t^{\mathrm{e,bat}} := b_t^{\mathrm{e}}-b_t^{\mathrm{e,pv}}$
to the feasible range determined by the remaining charging/discharging
energy margins and power limits:
\begin{align}
 b_t^\mathrm{e,bat}
 =
 \begin{cases}
  \min\left\{ \tilde b_t^\mathrm{e,bat}, B_t^\mathrm{dis} \right\}, &  \tilde b_t^\mathrm{e,bat} > 0, \\[4pt]
  \max\left\{ \tilde b_t^\mathrm{e,bat}, - B_t^\mathrm{chg} \right\}, & \tilde b_t^{\mathrm{e,bat}} \le 0 .
 \end{cases}    
\end{align}
With this formulation, no separate imbalance occurs for the battery, and $E_t^\mathrm{chgE}$ and $E_t^\mathrm{disE}$ in Eqs.\,\eqref{eq:chgE} and \eqref{eq:disE} are replaced with $\max\{ -b_t^\mathrm{e,bat}\Delta t,0\}$ and $\max\{b_t^\mathrm{e,bat}\Delta t, 0\}$, respectively. 
Furthermore, $E_t^\mathrm{chgAS}$ in \eref{eq:chgAS} is replaced with 
\begin{align}
  E_t^\mathrm{chgAS} = b_t^\mathrm{dn,bat}h_t^\mathrm{dn}. 
  \label{eq:chgAS_colocated}
\end{align}



\bibliographystyle{elsarticle-num}

\bibliography{cas-refs}



\end{document}